\documentclass[aoas,preprint,11pt]{imsart}

\RequirePackage[OT1]{fontenc}
\RequirePackage{amsthm,amsmath,amsfonts, epsfig}
\RequirePackage[numbers]{natbib}
\RequirePackage[colorlinks,citecolor=blue,urlcolor=blue]{hyperref}
\RequirePackage{hypernat}
 \usepackage[section] {placeins}

\newtheorem{theorem}{Theorem}[section]
\newtheorem{proposition}{Proposition}[section]
\newtheorem{lemma}{Lemma}[section]

\newtheorem{Remark}{Remark}[section]

\numberwithin{equation}{section}
\startlocaldefs
\numberwithin{equation}{section}
\theoremstyle{plain}

\endlocaldefs

\begin{document}

\begin{frontmatter}
\title{Asymptotically optimal tests when parameters are estimated}
\runtitle{Asymptotically optimal tests}

\begin{aug}
\author{\fnms{Tewfik} \snm{Lounis}
\ead[label=e1]{tewfik.lounis@gmail.com}} \runauthor{Lounis Tewfik}

\affiliation{
Universit\'e de
Caen-France}

\address{
Laboratoire de
math\'ematiques Nicolas Oresme\\CNRS UMR  6139 Universit\'e de
Caen\\ E-mail: \printead*{e1}}

\end{aug}

\begin{abstract}
The main purpose of this paper is to provide an asymptotically optimal test. The proposed statistic
 is of Neyman-Pearson-type when the parameters are estimated with a particular kind of estimators. It is shown
 that the proposed   estimators enable us to achieve this end.
Two particular cases, AR(1) and ARCH models were studied and the asymptotic power function was derived.
\end{abstract}


\end{frontmatter}

\section{Introduction}
Local asymptotic normality (LAN) for the log likelihood ratio was
studied for a several classes of nonlinear time series model, from
a LAN
 the contiguity property follows, for more details the interested reader may refer to \cite{Bickel1982},
  \cite{Swensen1985}, 
 and \cite{Hallin2008}. Applying the
contiguity property, we construct a statistic for testing a null
hypothesis $H_0$ against  the alternative hypothesis $H^{(n)}_1$,
often a various classical test statistics depends on the central
sequence which appears in the expression of the log likelihood
ratio, in the case when the parameter of the time series model is
known we obtain good properties of the test, precisely, the
optimality, see for instance \cite[Theorem 3]{HB}. However, in a
general case, particularly in practice, the parameter is
unspecified, in the expression of the estimate central sequence
appears an additional term which is non degenerate asymptotically.
The latter, alters  the power function of the constructed test.\\
In order to solve this very problem, and on a basis  of an
estimator of the unknown parameter, we introduce and define
another estimator which does not effects asymptotically  the power
function of the test, more precisely the additional term is
absorbed. The principle of this construction is to modify  one of
the component of the first estimator in order to avoid the
additional term, the details
of this method are expanded further in the section $2.$\\
The main purpose of this paper is to investigate the problem of
testing two hypothesis corresponding to  a stochastic model  which
is
 described in the following way. Let $\{(Y_{i},X_{i})\}$ be  a sequence  of stationary
and ergodic random vectors with finite second moment such that for
all $i \in \mathbb{Z}$, where $Y_{i}$ is a univariate random variable and
$X_{i}$ is a \emph{d}-variate random vector. We consider the class of
stochastic models
\begin{eqnarray}
    Y_i =T(Z_i) + V(Z_i)\,\epsilon_i, \quad
    i\in\mathbb{Z},\label{modelprincipal}
\end{eqnarray}
\noindent where, for given non negative  integers $q$  and $s$,
the random vectors $ Z_i$  is equal to $(Y_{i-1}, Y_{i-2}, \ldots,
Y_{i-s},X_i,X_{i-1},\ldots,X_{i-q})$, the $\epsilon_{i}$'s are
centered i.i.d. random variables with unit variance and density
function $f(\cdot)$, such that for each $i \in \mathbb{Z}$,
$\epsilon_{i}$ is independent of the filtration $\mathcal{F}_{i} =
\sigma(Z_j, j \leq i),$ the real-valued functions $T (\cdot)$ and
$V (\cdot)$ are assumed to be unknown. We consider the problem of
 testing whether the bivariate vector of
functions $(T (\cdot),V (\cdot))$ belongs to a given class of
parametric functions or not. More precisely, let
$$\mathcal{M}=\left\{\left(m({\rho},\cdot), \sigma({\theta},\cdot)\right)
,~ ({\rho}^\prime,{\theta}^\prime)^\prime \in \Theta_1 \times
\Theta_2 \right\},
$$ $ \Theta_1 \times
\Theta_2\subset\mathbb{R}^\ell\times\mathbb{R}^p,$
 $\mathring{\Theta}_1\neq \emptyset,$    $\mathring{\Theta}_2\neq \emptyset,$
where for all set $A$, $\mathring{A}$ denotes the interior of the
set $A$ and the script ``$~~^\prime~~ $'' denotes the transpose,
$\ell$ and $p$ are two positive integers, and each one of the two
functions $m({\rho},\cdot)$ and $\sigma({\theta},\cdot)$ has a
known form such that $\sigma({\theta},\cdot)>0.$ \noindent For a
sample of size $n$, we derive a test of \begin{eqnarray}H_0:
\left[(T(\cdot),V(\cdot))\in \mathcal{M}\right]
\mbox{~~~against~~~}
H_1:\left[(T(\cdot),V(\cdot))\notin\mathcal{M}\right].\end{eqnarray}
It is easy to see that the null hypothesis $H_0$ is equivalent to
\begin{eqnarray}
  H_0:[(T(\cdot),V(\cdot)] &=&
  \Big(m({\rho}_{0},\cdot),\sigma({\theta}_0,\cdot)\Big),\label{principal null hypothesis}
\end{eqnarray}
 while the alternative hypothesis $H_1$ is equivalent to
\begin{eqnarray*}
H_1:[(T(\cdot),V(\cdot)] &\neq
&\Big(m({\rho}_0,\cdot),\sigma({\theta}_0,\cdot)\Big),
\end{eqnarray*}for some $({\rho}_{0}^\prime,{\theta}_{0}^\prime)^\prime \in
\Theta_1 \times \Theta_2$. \\In the sequel, our study will be
focused on the following  alternative hypotheses. For all integers
$n\geq 1$, the alternative hypothesis $H^{(n)}_1$ is defined by
the following equation
\begin{eqnarray}
H^{(n)}_1:[(T(\cdot),V(\cdot)]&=& \Big(m({\rho}_0,\cdot)
+\,n^{-\frac{1}{2}}G(\cdot),\sigma({\theta}_0,\cdot)+\,
n^{-\frac{1}{2}}S(\cdot)\Big),\label{principal alternative
hypothesis}
\end{eqnarray}
where $G(\cdot)$ and $S(\cdot)$ are two specified real functions.
The situation is different in the case when the used statistic is
the Neyman-Pearson test which is based on the log-likelihood ratio
 $\Lambda_n$ defined as follows
\begin{equation}
   \Lambda_n=\log\left(\frac{f_n}{f_{n,0}}\right)=\sum_{i=1}^{n}\log(g_{n,i}),\label{logarithm}
\end{equation}
where  $f_{n,0}(\cdot)$ and $f_{n}(\cdot)$  denote the probability densities of the
random vector $(Y_1,\ldots,Y_n)$ corresponding to the null
hypothesis and the alternative hypothesis, respectively.\\
The use of the Neyman-Pearson statistics needs to resort to
the following conditions:\\
Under the hypothesis $H_0$, there exists a random variable
$\mathcal{V}_n$ such that $$\mathcal{V}_n
\stackrel{\mathcal{D}}{\longrightarrow} \mathcal{N}(0,{\tau}^2),$$
where $\stackrel{\mathcal{D}}{\longrightarrow}$ denotes the
convergence in distribution and some constant $\tau>0$ depending
on the parameter
$\phi_0=(\rho_{0}^\prime,\theta_{0}^\prime)^\prime,$ such that
\begin{eqnarray}
 \Lambda_n =\mathcal{V}_n(\phi_0) -\frac{{\tau}^2(\phi_0)}{2} + o_{P}(1)\label{lan}.
 \end{eqnarray}
The equality (\ref{lan}) is a modified version of the LAN given by
\cite[Theorem 1]{HB}. We mention that there exist other versions
of the LAN, we may refer to \cite{l}, \cite{HM}, and the
references therein. On the basis of the LAN, an efficient test of
linearity based  on Neyman-Pearson-type statistics was obtained in
a class of nonlinear time series models contiguous to a
first-order autoregressive process $AR(1)$ and its asymptotic
power function is derived (see, \cite[Theorem 1 and Theorem
3]{HB}).The expression of the obtained test depends on the central
sequence $\mathcal{V}_n(\phi_0)$ which itself depends  on the
parameter $\phi_0$. In a general case the parameter $\phi_0$ is
unspecified, so, in order to estimate it, we introduce, under some
assumptions, an estimate preserving, asymptotically, the power on
Neyman-Pearson test  when we replace, in the expression of the
statistics, the parameter $\phi_0$ by an appropriate  estimator,
 $\bar{\phi}_n.$ Say,  this estimator  will be constructed on the
 tangent space with the direction of the partial derivatives
 of the central sequences in  $\hat{\phi}_n$, where $\hat{\phi}_{n}$ is a $\sqrt{n}$-consistent estimator of $\phi_0$.
 In the sequel, $\bar{\phi}_n$  will be called a \emph{modified estimate} (M.E.). \\
 This paper is organized as follows: Section \ref{section2}  describes  the methodology
 used to construct  the M.E. In Section \ref{section3},
we give the asymptotic properties of the proposed estimate. In
Section \ref{section4}, we conduct a simulation in order to
evaluate the power of the proposed test. All mathematical
developments are relegated to the Section $\ref{proof}$.

\section{Estimation with modifying one component }\label{section2}
Consider the problem of testing the two hypothesis $H_0$ against
$H^{(n)}_1$ which are given in (\ref{principal null hypothesis})
and (\ref{principal alternative hypothesis}) respectively and
corresponding to the stochastic model (\ref{modelprincipal}). We
assume that the LAN (\ref{lan}) of the
model (\ref{modelprincipal}) is established, for example refer to \cite{HB}.\\
 \noindent Let $\hat{\phi}_{n} =({\hat{\rho}_{n}}^\prime
,{\hat{\theta}_{n}}^\prime)^\prime $ a $\sqrt{n}$-consistent
estimate of the parameter
$\phi_0=(\rho_{0}^\prime,\theta_{0}^\prime)^\prime,$
where\begin{eqnarray*}
{\hat{\rho}_{n}}^\prime&=&\Big(\hat{\rho}_{n,1},\dots,\hat{\rho}_{n,\ell}\Big)
\mbox{,}\quad
{\hat{\theta}_{n}}^\prime=\Big(\hat{\theta}_{n,1},\dots,\hat{\theta}_{n,p}\Big),\\
{\rho_{0}}^\prime&=&\Big(\rho_{1},\dots,\rho_{\ell}\Big)\quad
\mbox{and} \quad
{\theta_{0}}^\prime=\Big(\theta_{1},\dots,\theta_{p}\Big).
\end{eqnarray*}Our purpose is to construct another estimate ${\bar{\phi}_{n}}^\prime$ of
the parameter $(\rho^\prime_{0};\theta^\prime_{0})^\prime,$ such
that the following fundamental equality is
fulfilled\begin{eqnarray}\label{eqaaa1}
 \mathcal{V}_n(\bar{\phi}_{n}) - \mathcal{V}_n(\hat{\phi}_{n})&=&   D_n,
\end{eqnarray} where  $D_n$ is a specified bounded random function. In the sequel, the
 functions\\
$(\rho,\cdot)\rightarrow m(\rho,\cdot)$ and
$(\theta,\cdot)\rightarrow \sigma(\theta,\cdot)$ are assumed to be
twice  differentiable. Our goal, is to find an estimate
$\bar{\phi}_{n}$ satisfying (\ref{eqaaa1}) pertaining to the
tangent space $\Gamma_{n}$, such that, for $(X^\prime,
Y^\prime)^\prime \in \mathbb{R^\ell}\times {\mathbb{R}}^p$, the
following equation holds
\begin{eqnarray*} \Gamma_{n}: \mathcal{V}_n
((X, Y)) - \mathcal{V}_n(\hat{\phi}_{n})
&=&\partial\mathcal{V}_n^\prime(\hat{\phi}_{n}).\Big({(X -
{\hat{\rho}_{n}})}^\prime, {(Y
-\hat{\theta}_{n})}^\prime\Big)^\prime,
\end{eqnarray*}
where\begin{eqnarray*}
\partial\mathcal{V}_n(\hat{\phi}_{n})^\prime&=&\Big(\frac{\partial\mathcal{V}_n(\hat{\phi}_{n})}{\partial
\rho_{1}},\ldots,\frac{\partial\mathcal{V}_n(\hat{\phi}_{n})}{\partial
\rho_{\ell}},\frac{\partial\mathcal{V}_n(\hat{\phi}_{n})}{\partial
\theta_{1}},\ldots,
\frac{\partial\mathcal{V}_n(\hat{\phi}_{n})}{\partial
\theta_{p}}\Big),
\end{eqnarray*} and the script $"\cdot"$ denotes the inner
product.\\
 With the connection with the equality (\ref{eqaaa1}), the  new estimate is then given by imposing  that the value $(X^\prime,
 Y^\prime)^\prime$ satisfied the following identity\begin{eqnarray}
  D_n&=&\partial\mathcal{V}_n(\hat{\phi}_{n})^\prime.\Big({(X - {\hat{\rho}_{n}})}^\prime, {(Y -\hat{\theta}_{n})}^\prime\Big
)^\prime.\label{contrainte}
\end{eqnarray}
\noindent  Clearly,  the equation (\ref{contrainte}) has $\ell +
p$ unknown values, so it has an infinity of solutions, after
modification of the $j_{n}$-th  component of the first estimate
$\hat{\rho}_{n}$, we shall propose an element in tangent space
$\Gamma_{n}$ which satisfies the equality (\ref{contrainte}). We
obtain then a new estimate
${\bar{\phi}_n}^\prime={\phi_{n}^{(1,j_{n})}}^\prime=({{\bar{\rho}}^\prime_{n}}
,{\hat{\theta}^\prime_{n}})^\prime$ of the unknown parameter
$\phi_0$, where\begin{eqnarray*}
{\bar{\rho_{n}}}^\prime&=&\Big(\bar{\rho}_{n,1},\dots,\bar{\rho}_{n,\ell}\Big),
\end{eqnarray*} and  such that:  for s $\in \{1,\ldots,\ell\}$,
$\bar{\rho}_{n,s}=\hat{{\rho}}_{n,s} ~\mbox{if}~s\neq j_{n}~\mbox{
and }~\bar{\rho}_{n,j_{n}}\neq\hat{{\rho}}_{n,j_{n}}. $\\
The use of the  notation $\phi_{n}^{(1,j_{n})}$ explains that we
obtain the new estimate $\bar{\phi}_n$ of the parameter $\phi_0$
when we change  in  the expression of the estimate $ \hat{\phi}_n$
the $j_{n}$ component with respect to the first estimate
$\hat{\rho}_n$ corresponding to the step $n$ of the estimation. It
follows from the equality (\ref{eqaaa1}) combined with the
constraint (\ref{contrainte}) that \begin{eqnarray}
 \mathcal{V}_n(\phi_{n}^{(1,j_{n})})- \mathcal{V}_n(\hat{\phi} _{n})
 &=&\sum _{s=1}^{\ell}\frac{\partial\mathcal{V}_n(\hat{\phi}_{n})}{\partial\rho_{s}}(\bar{\rho}_{n,s} -
 \hat{{\rho}}_{n,s}) + \sum _{t=1}^{p}\frac{\partial\mathcal{V}_n(\hat{\phi}_{n})}{\partial\theta_{t}}(\bar{\theta}_{n,t} -
 \hat{{\theta}}_{n,t}),\nonumber \\
  &=&\frac{\partial\mathcal{V}_n(\hat{\phi}_{n})}{\partial\rho_{j_{n}}}(\bar{\rho}_{n,j_{n}} -
 \hat{{\rho}}_{n,j_{n}}).
\end{eqnarray}
 By imposing the following condition\begin{eqnarray}
  \frac{\partial {\mathcal{V}}_n(\hat{\phi}_{n})}{\partial\rho_{j_{n}}}
  &\neq&0,
  \label{gradient 1}
\end{eqnarray}
\noindent and with the  use of the equality (\ref{contrainte})
combined with (\ref{gradient 1}), we deduce that
\begin{eqnarray}
  \bar{\rho}_{n,j_{n}} &=&
  \frac{D_n}{\frac{\partial\mathcal{V}_n(\hat{\phi}_{n})}{\partial\rho_{j_{n}}}}
  +\hat{{\rho}}_{n,j_{n}}.\label{perturbation1}
\end{eqnarray}
In summary, we define the modified  estimate by
$$\bar{\phi}_{n}^\prime={\phi_{n}^{(1,j_{n})}}^\prime=\Big(\hat{\rho}_{n,1},\dots,\hat{\rho}_{n,j_{n}-1},
\bar{\rho}_{n,j_{n}},\hat{\rho}_{n,j_{n}+1},\dots,\hat{\rho}_{n,\ell}
,\hat{\theta}_{n,1},\dots,\hat{\theta}_{n,p}\Big)^\prime.$$
\noindent With a same reasoning as the previous case and after
modifying the $k_{n}$-th component with respect to the second
estimate, we shall define a new estimate
$$\bar{\phi_{n}}^\prime={\phi_{n}^{(2,k_{n})}}^\prime=({{\hat{\rho}}_{n}}^\prime,{\bar{\theta}_{n}}^\prime)^\prime,$$ such that
for $t \in \{1,\ldots,p\}$
$$\bar{\theta}_{n,t}=\hat{\theta}_{n,t} ~ \mbox{if}~t\neq
k_{n}~\mbox{ and
}~\bar{\theta}_{n,k_{n}}\neq\hat{\theta}_{n,k_{n}}.$$
 we obtain
\begin{eqnarray}
 \mathcal{V}_n(\phi_{n}^{(2,k_{n})})- \mathcal{V}_n(\hat{\phi}_{n}) &=&
 \frac{\partial\mathcal{V}_n(\hat{\phi}_n)}{\partial\theta_{k_{n}}}(\bar{\theta}_{n,k_{n}} -
 \hat{\theta}_{n,k_{n}}).
\end{eqnarray}
Under the following condition\begin{eqnarray}
  \frac{\partial\mathcal{V}_n(\hat{\phi}_{n})}{\partial\theta_{k_{n}}} &\neq& 0,\label{gradient 2}
\end{eqnarray}
\noindent it follows  from the equality (\ref{contrainte})
combined with (\ref{gradient 2}), that
\begin{eqnarray}
  \bar{\theta}_{n,k_{n}} &=&
  \frac{D_n}{\frac{\partial\mathcal{V}_n(\hat{\phi}_{n})}{\partial\theta_{k_{n}}}}
  +\hat{{\theta}}_{n,k_{n}}.\label{perturbation2}
\end{eqnarray}
In summary, we obtain the modified  estimate
$$\bar{\phi}_{n}^\prime={\phi_{n}^{(2,k_{n})}}^\prime=\Big(\hat{\rho}_{n,1},\dots,\hat{\rho}_{n,\ell}
,\hat{\theta}_{n,1},\dots,\hat{\theta}_{n,k_{n}-1},\bar{\theta}_{n,k_{n}},\hat{\theta}_{n,k_{n}+1},\dots,\hat{\theta}_{n,p}\Big)^\prime.$$
The estimate $\phi_{n}^{(1,j_{n})}$ (respectively,
$\phi_{n}^{(2,k_{n})}$) is called a modified estimate in
$j_{n}$-th component with respect to the first
 estimate (respectively, in $k_{n}$-th component with respect to second estimate),
we denote this estimate by (M.E.).
\begin{Remark}
 For each step $n$ of the estimation  corresponding a value of the
 position
 $j_{n}$ or $k_{n}$ of the component where the estimate was modified.
\end{Remark}
\section{Properties of the (M.E.)}\label{section3}
\subsection{Consistency}
Throughout, $\hat{\phi}_{n}$ is a $\sqrt{n}$-consistent estimate
of the unknown parameter $\phi_0.$
 \noindent The conditions
(\ref{gradient 1}) and  (\ref{gradient 2}) are not sufficient to
get the consistency of the modified estimate (M.E.). In order to
get its consistency, we need  to resort to  one of  the following
additional conditions .
\begin{enumerate}
\item[(C.1)]
$$ \frac{1}{\sqrt{n}} \frac{\partial
{\mathcal{V}}_n(\hat{\phi}_{n})}{\partial\rho_{j_{n}}}
  \stackrel{P}{\longrightarrow}c_{1} \quad\mbox{as}\quad n\rightarrow\infty,
  \label{Consisatncegradient 1}$$
\item[(C.2)]
$$
\frac{1}{\sqrt{n}}\frac{\partial\mathcal{V}_n(\hat{\phi}_{n})}{\partial\theta_{k_{n}}}\stackrel{P}{\longrightarrow}c_{2}
\quad\mbox{as}\quad n\rightarrow\infty
,\label{Consistancegradient2}
$$
where $c_{1}$ and $c_{2}$ are two constantes, such that
$c_{1}\neq0$ and $c_{2}\neq0.$
\end{enumerate}
\noindent Our first result concerning the consistency of the
proposed estimate is summarized in the following proposition.
\begin{proposition}\label{consistence}
Under (\ref{gradient 1}) and $(C.1)$ ((\ref{gradient 2}) and
$(C.2)$, respectively), the estimate $\phi_{n}^{(1,j_{n})}$
($\phi_{n}^{(2,k_{n})}$, respectively) is a $\sqrt{n}$-consistent
estimator of the unknown parameter $\phi_0$.
\end{proposition}
  In practice, it is not easy to verify  the condition
  $(C.1)$ (respectively, $(C.2)$), in the case when
  the unknown parameter $\phi_0$ is univariate, a sufficient condition will be
  stated in Lemma (\ref{sufficientcondition}), in this case, we need the
following assumption:
\begin{enumerate}
 \item[(C.3)]: For all real sequence ${(\eta_n)}_{n\geq1}$ with
 values in the interval $ [0,1]$, we have:
$$ \frac{1}{\sqrt{n}} \mathcal{\ddot{V}}_n(\eta_n \phi_{0} +
(1-\eta_n)\hat{\phi}_n)) =O_{P}(1),$$\label{ConsistanceDERIVAIVE1}\\
where $\mathcal{\ddot{V}}_n$ is a second derivative of
$\mathcal{V}_n.$

\end{enumerate}
\begin{Remark}
In a problem of testing the two hypothesis $H_0$ against
$H^{(n)}_1,$ and when the error  $\epsilon_{i}$'s are centered
i.i.d. and $\epsilon_{0} \stackrel{\mathcal{D}}{\longrightarrow}
\mathcal{N}(0,1),$  a  large classe  of  time series model
satisfied the condition $(C.3)$, for instance, we cite the
nonlinear time series contiguous to AR(1) processes , the details
are expanded further  later in  the proofs of the Propositions
(\ref{AR1ECARTfirst}) and (\ref{ARCHECARTsecond}) .
\end{Remark}
Now, we may state the sufficient condition which implies
assumptions  $(C.1)$  corresponding to the case when the parameter
of the time series model is
univariate.\begin{lemma}\label{sufficientcondition} Let
$\hat{\phi}_n$ be a $\sqrt{n}$-consistent estimate of the
 parameter $\phi_0$. Let $c_1$  be a constant, such that $c_1\neq0$,  then we
 have:
\begin{itemize}
    \item[(i)]  Under $(C.3)$, if  $\frac{1}{\sqrt{n}}
{\mathcal{\dot{V}}}_n(\phi_0)
  \stackrel{P}{\longrightarrow}c_{1}$, as $
  n\rightarrow\infty,$ then $\forall A>0,$ $$P\left(\left|\frac{1}{\sqrt{n}}{\mathcal{\dot{V}}}_n(\hat{\phi}_n)-c_1\right|>A\right)\rightarrow0,
~as~
  n\rightarrow\infty.$$
 \end{itemize}
  \end{lemma}
\subsection{Absorbtion of the error}
\noindent Consequently, with the modified estimate and in the case
when the error between two central sequences is bounded, it is
 possible to absorb this error, this result is stated and proved
 in the following proposition.
\begin{proposition}\label{M.E} Let
${\hat{\phi}_n}^\prime$ be an estimate ($\sqrt{n}$ consitency) of
the parameter ${(\rho^\prime,\theta^\prime)}^\prime$. We assume
that there exists a known bounded function $D_n$, such
that\begin{eqnarray}
 \mathcal{V}_n (\hat{\phi}_{n}) &=& \mathcal{V}_n(\phi_0)-
 D_n+o_{P}(1)\label{link between the central sequences}.
\end{eqnarray}Then,  there exists an estimate ${\bar{\phi_{n}}}^\prime$ of
$(\rho^\prime,\theta^\prime)^\prime$ such that
\begin{eqnarray*}
 \mathcal{V}_n(\bar{\phi_{n}}) &=&
 \mathcal{V}_n(\phi_0)+o_{P}(1).
\end{eqnarray*}
\end{proposition}
\begin{Remark}
The equality (\ref{link between the central sequences}) gives the
link between the estimated central sequences  $\mathcal{V}_n
(\hat{\phi}_{n})$ and the central sequence
$\mathcal{V}_n(\phi_0)$. Sometimes it is not easy to establish the
form of the function $D_n,$ in the next section, we propose, under
some assumptions, how to specify this function in the two cases,
i.e.,  the case when the problem of testing the linearity and
nonlinearity of the $s$-th order  and  the case of time series
model with conditional heteroscedasticity  respectively
corresponding to the equalities\begin{eqnarray*}
 Y_i =\rho_0 Y_{i -1} + \alpha G(Y(i -1))+\epsilon_i
 \mbox{~~and~~} \\Y_i =\rho_0 Y_{i -1} + \alpha G(Y(i -1)) +\sqrt{1+ \beta B(Y(i -1))} \,\epsilon_i,
\end{eqnarray*}respectively, where   $ Y(i -1) = \Big(Y_{i-1}, Y_{i-2},
\ldots,Y_{i-s}\Big),$ $\alpha$ and $\beta$ are  real parameters
and the $\epsilon_{i}$'s are centered i.i.d. random variables with
unit variance and density function $f(\cdot)$.
\end{Remark}
 Throughout, we assume that the function $f(\cdot)$ is positive
with a third  derivative, we denote by $\dot{f}(\cdot)$,
$\ddot{f}(\cdot)$  and  $f^{(3)}(\cdot)$
 the first, the second and the third derivative respectively. For all $x \in
\mathbb{R},$ let
\begin{equation*} M_{f}(x)=\frac{\dot{f}(x)}{f(x)}.
\end{equation*} According to the notation (\ref{logarithm}), we suppose that the three following conditions are satisfied :
\begin{itemize}
    \item (L.1): $ \max_{1\leq i\leq n}| g_{n,i} - 1|=o_{P}(1),$
    \item (L.2): there exists a positive constante ${\tau}^2$ such that\\
   $ \sum_{i=1}^{n}(g_{n,i} - 1)^2 = {\tau}^2 + o_{P}(1),$
   \item (L.3):  there exists a-$\mathcal{F}_{n}$  mesurable
   $\mathcal{V}_n$ satisfying $ \sum_{i=1}^{n}(g_{n,i} - 1) = \mathcal{V}_n +
   o_{P}(1),$ where $\mathcal{V}_n\stackrel{\mathcal{D}}{\longrightarrow}
   \mathcal{N}(0,{\tau}^2).$
\end{itemize}
Conditions $(L.1),$ $(L.2)$ and $(L.3)$ imply under $H_0$ the
local asymptotic normality LAN corresponding to  the equality
(\ref{lan}), for more details see (\cite[Theorem 1]{HB}). This
last theorem is the fundamental tool used later to aim to
establish the LAN for the considering models.

\subsection{Link between central sequences in nonlinear time series contiguous to AR(1)
processes}\label{firstproblem}

 \noindent Consider the $s$-th order (nonlinear) time
series
\begin{eqnarray}
    Y_i =\rho_{0} Y_{i -1} + \alpha \,G(Y(i -1)) + \epsilon_i\mbox{,} \quad |\rho_{0}| <1.\label{FIRSTAR1MODEL}
\end{eqnarray}
  In this case and with the comparison to the
equality  (\ref{modelprincipal}), we have
\begin{eqnarray*}
 Z_i=Y_i \quad \mbox{,}\quad  T(Z_i)=\rho_{0} Y_{i -1} + \alpha \,G(Y(i -1)) \quad \mbox{and}\quad
V(Z_i)=1.
\end{eqnarray*}
\noindent In the sequel, it will be assumed that  the model
\label{principal nonlinear} is a stationary and ergodic  time
series with finite second moment. We consider the problem of
testing the null hypothesis $H_0:\alpha=0$ against the
alternative hypothesis $H^{(n)}_1:\alpha =n^{-\frac{1}{2}},$ with
the comparison  to (\ref{principal null hypothesis}) and
(\ref{principal alternative hypothesis}), we have
\begin{eqnarray*}
\Big(m({\rho}_{0},Y_{i-1}),\sigma({\theta}_0,Y_{i-1})\Big)^\prime&=&\Big(
\rho_0 \,Y_{i-1},1\Big)^\prime \mbox{,}\quad
\mathcal{M}=\left\{m({\rho},\cdot), {\rho} \in \Theta_1  \right\}
\mbox{,}\\ {Z_i}^\prime &=& \Big(Y_{i-1},\dots, Y_{i-s}\Big) \quad
\mbox{and}\quad S(\cdot)=0.
\end{eqnarray*}
Note that this problem of testing is equivalent to test the
linearity of the $s$-th $AR(1)$  time series model  when
($\alpha=0$) against the nonlinearity of the  $s$-th $AR(1)$  time
series model when ($\alpha=n^{-\frac{1}{2}}$). \\
Throughout, the scripts "$\|\cdot\|_{{\ell}+p}$ " ,
"$\|\cdot\|_{{\ell}}$ " and "$\|\cdot\|_p$ " denote the euclidian
norms in $\mathbb{R}^{{\ell}+p}$, $\mathbb{R}^{{\ell}}$ and
$\mathbb{R}^{p}$ respectively. It will be assumed that the
conditions (A.1) and (A.2) are satisfied, where
\begin{itemize}
    \item (A.1): There exists positive constants $\eta$ and $c$ such that for all $u$ with
    $\|u\|_{{\ell}+p}>\eta$, $G(u)\leq c\|u\|_{{\ell}+p}$.
   \item (A.2): for a location family $\{f(\epsilon_i -c),~ -\infty
    <c<-\infty\}$, there exist a square integrable functions
    $\Psi_1,$ $\Psi_2$ and a constant $\delta$ such that for all $\epsilon_i$ and
    $|c|<\delta,$ such that :\begin{eqnarray*}
      \Big|\frac{{d}^k{f(\epsilon_i -c)}}{f(\epsilon_i)\,dc^k\,}\Big|
      \leq\Psi_k(\epsilon_i)\mbox{,}\quad \mbox{for}\quad k=1,2.
    \end{eqnarray*}
\end{itemize}
Under the conditions (A.1) and (A.2) the LAN of the time series
model (\ref{FIRSTAR1MODEL}) was established in (\cite[Theorem
2]{HB}), the proposed test $T_{n}$ is the Neyman-Pearson statistic
which is  given by the following equality\begin{eqnarray*}
  T_{n} &=& I{\left\{{\frac{\mathcal{V}_{n}(\rho_0)}{\tau(\rho_0)}\geq Z(\alpha)}\right\}},\label{test}
\mbox{~~~where~~~}\tau^2=\mathbf{E}({M^2_{f}}(\epsilon_{0}))\mathbf{E}(G^2(Y(0))),
\end{eqnarray*}and $Z(\alpha)$ is the $(1 -\alpha)$-quantile of a standard normal
distribution $\Phi(\cdot).$ In this case, the central sequence is
given by the following equality\begin{eqnarray*}
\mathcal{V}_{n}(\rho_0)=-\frac{1}{\sqrt{n}}\sum_{i=1}^{n}M_{f}(\epsilon_i)G(Y(i-1)),
\mbox{~~~where~~~}\tau^2=\mathbf{E}({M^2_{f}}(\epsilon_{0}))\mathbf{E}(G^2(Y(0))),
\end{eqnarray*} and such that under $H_0,$
$\mathcal{V}_{n}(\rho_0)\stackrel{\mathcal{D}}{\longrightarrow}\mathcal{N}(0,
{\tau}^2).$ The asymptotic power of the test is derived and equal
to $1 - \Phi(Z(\alpha)-{\tau}^2 )$, recall that when $\rho_0$ is
known, this test is asymptotically optimal, for more
details see \cite[Theorem 3]{HB}.\\
 \noindent Our aim is to  specify the form of the function $D_n$
 which is defined in (\ref{link between the central sequences}),
 the parameter $\rho_0$ is estimated by  the $\sqrt{n}$-consistent estimator 
$\hat{\rho}_n$ and the residual $\epsilon_i$  is estimated by
$\hat{\epsilon}_{i,n} = Y_i  - Y_{i-1} \hat{\rho}_n.$  We have the
following statement: \begin{proposition}\label{AR1ECARTfirst}
Assume that the conditions $(A.1)$ and $(A.2)$ hold and
$\epsilon_{i}$'s are centered i.i.d. and $\epsilon_{0}
\stackrel{\mathcal{D}}{\longrightarrow} \mathcal{N}(0,1)$. We have
\begin{eqnarray}
 \mathcal{V} (\hat{\rho}_n) = \mathcal{V}_n(\rho_0)-D_n+o_{P}(1),\label{link between the central sequences in AR1}
\end{eqnarray} where\begin{eqnarray}
 D_n= -c_1\sqrt{n}(\hat{\rho}_n
- \rho_0) \label{specificationD},\\
\bar{\rho}_{n} = \frac{D_n}{{\mathcal{\dot{V}}_n(\phi_{n})}}
  +\hat{{\rho}}_{n}\mbox{~~and~~}
c_1=-\mathbb{E}\Big[Y_{0}G(Y(0))\Big]\label{specificationme}.
\end{eqnarray}
\end{proposition}
\begin{Remark}
\begin{itemize}
    \item The use of the ergodicity of the model imposes to require the
condition  $\mathbb{E}\Big[Y_{-1}G(Y_0)\Big]<\infty,$ therefore we
choose the function $G(\cdot)$ in order to get this condition. For
instance, we shall choose $G(Y(i-1))=\frac{2a}{1+Y^2_{i-1}},$
where $a\neq0.$
    \item With  this choice of the function $G$, the condition
    $(A.1)$ remains satisfied, in fact, we can  remark that $|G(u)|\leq2|a|,$
    then for all u with $\|u\|_{{\ell}+p}\geq \eta $ we have $G(u)\leq 2a \times \|u\|_{{\ell}+p}\times
    \frac{1}{\|u\|_{{\ell}+p}}\leq \frac{2a}{\eta}\times  \|u\|_{{\ell}+p},$ therefore, we
    shall choose $c=\frac{2a}{\eta}.$
\end{itemize}
\end{Remark}
\subsection{ An extension to ARCH processes}\label{secondmodel}
\noindent Consider the following time series model with
conditional heteroscedasticity
\begin{eqnarray}
    Y_i =\rho_0 Y_{i -1} + \alpha \,G(Y(i -1)) + \sqrt{1  + \beta B(Y(i -1))} \,\epsilon_i, \quad
    i\in\mathbb{Z}.\label{model with conditional
heteroscedasticity}
\end{eqnarray}
 It is assumed that the model (\ref{model with
conditional heteroscedasticity}) is ergodic and stationary. It
will be assumed that the conditions (B.1), (B.2) and (B.3)
are satisfied, where
\begin{itemize}
    \item (B.1): The fourth order moment of the stationary distributions of (\ref{model with conditional heteroscedasticity})  exists.
    \item (B.2):  There exists a positive constants $\eta$ and $c$ such that for all $u$
    with\\
    $\|u\|_{{\ell}+p}>\eta$,~~~ $B(u)\leq c\|u\|_{{\ell}+p}^2.$
   \item (B.3): for a location family $\{b^{-1}f(\frac{\epsilon_i -a}{b}),~ -\infty
    <a<-\infty, ~b>0\}$, there exists a square integrable
function $\varphi(\cdot)$, and a strictly
    positive real $\varsigma$, where $\varsigma>\max(|a|,|b-1|)$, such that,
  $$\left|\frac{{\partial}^2{b^{-1}f\left(\frac{\epsilon_i -a}{b}\right)}}{f(\epsilon_i)\,\partial
                            a^j\,
                            \partial b^k}\right|\leq\varphi(\epsilon_i),$$
where $j$ and $k$ are two positive integers such that
                $j + k = 2.$
 \end{itemize}
 We consider  the  problem of testing the
null hypothesis $H_0$ against the alternative hypothesis
$H^{(n)}_{1}$ such that
\begin{eqnarray*}
 H_0 &:& m(\rho, Z_i) ={\rho}_0 Y_{i-1}
\quad \mbox{and} \quad  \sigma({\theta}_0,\cdot)=1,\\
 H^{(n)}_{1} &:&m(\rho, Z_i) ={\rho}_0 Y_{i-1}+\,{n}^{-\frac{1}{2}}G(Y(i -1))\mbox{~~and~~ } \sigma({\theta}_0,Z_i)=\sqrt{1
  + n^{-\frac{1}{2}} B(Y(i -1))}.
\end{eqnarray*}
Remark that $H_0$, $H^{(n)}_{1}$ correspond to $\alpha= \beta=0$
(linearity of (\ref{model with conditional heteroscedasticity}))
and $\alpha=\beta={n}^{-\frac{1}{2}}$ (non linearity of
(\ref{model with conditional heteroscedasticity}))
 with the comparison to the
equality  (\ref{modelprincipal}), we have
\begin{eqnarray*}
 Z_i=Y_i\mbox{,}\quad  T(Z_i)=\rho_0 Y_{i -1} + \alpha \,G(Y(i -1))\mbox{~~~and~~~}V(Z_i)=\sqrt{1  + \beta B(Y(i -1))}.
\end{eqnarray*}
Note that when $n$ is large, we have
\begin{eqnarray*}
  \sigma({\theta}_0,Z_i)=\sqrt{1 + n^{-\frac{1}{2}} B(Y(i -1))}\sim 1
  + \frac{n^{-\frac{1}{2}}}{2} B(Y(i -1))=1
  +n^{-\frac{1}{2}} S(Y(i -1)).
\end{eqnarray*}
Under the conditions (A.1), (B.1), (B.2), and (B.3), the LAN was
established in \cite[Theorem 4]{HB}, an efficient test is obtained
and its power function is derived. In this case, the central
sequence is given by the following equality
\begin{eqnarray*}
\mathcal{V}_{n}(\rho_0)=-\frac{1}{\sqrt{n}}\left\{\sum_{i=1}^{n}M_{f}(\epsilon_i)G(Y(i-1))
+ \sum_{i=1}^{n}(1 + \epsilon_i M_{f}(\epsilon_i))B(Y(i
-1))\right\},
\end{eqnarray*}
 such that under $H_0,$ $$\mathcal{V}_{n}(\rho_0)\stackrel{\mathcal{D}}{\longrightarrow}\mathcal{N}(0,  {\tau}^2),$$
where
\begin{eqnarray*}
  {\tau}^2 = I_0 \mathbf{E}\left(G(Y(0)
        \right)^2
+ \frac{(I_2 - 1)}{4} \mathbf{E}\left(B(Y(0) \right)^2 + I_1
\mathbf{E}\left(G(Y(0))B(Y(0)\right)\\\mbox{~~where~~}
I_j=\mathbf{E}\Big({\epsilon}^j_{0}{M^2_{f}}(\epsilon_{0})\Big)
\mbox{~and~} j=0,1,2.
 \end{eqnarray*}
 The proposed test is then given by
  \begin{eqnarray}\label{lastexpresion}
  T_{n} &=& I{\left\{{\frac{\mathcal{V}_{n}(\rho_0)}{\tau(\rho_0)}\geq
  Z(\alpha)}\right\}}.
\end{eqnarray}
By the subsisting $\rho_0$ by its $\sqrt{n}$-consistent estimator
$\hat{\rho}_n$ in the expression of the central sequence, we shall
state the following proposition:
\begin{proposition}\label{ARCHECARTsecond}Suppose that the conditions $(A.1)$, $(B.1)$, $(B.2)$ and $(B.3)$ hold and
  $\epsilon_{i}$'s are centered
i.i.d. and $\epsilon_{0} \stackrel{\mathcal{D}}{\longrightarrow}
\mathcal{N}(0,1)$. We have
\begin{eqnarray}
\mathcal{V} (\hat{\rho}_n) &=& \mathcal{V}_n(\rho_0)-
 D_n+o_{P}(1),\label{link between the central sequences in ARCH}
\end{eqnarray}
where
\begin{eqnarray}
 D_n= -c_1\sqrt{n}(\hat{\rho}_n
- \rho_0) \label{specification2},\\
\bar{\rho}_{n} = \frac{D_n}{{\mathcal{\dot{V}}_n(\phi_{n})}}
  +\hat{{\rho}}_{n}\mbox{~~and~~} c_1=-\mathbb{E}\Big[Y_{0}G(Y(0))\Big].
\end{eqnarray}
\end{proposition}
\subsection{Optimality of the proposed test}
Throughout, $\bar{T}_n$ and $\bar{\tau}$ are the statistics test
and the constant   respectively obtained with the subsisting of
the unspecified parameter $\phi_0$ by its modified estimate
$\bar{\phi}_n$ in the expression of the test (\ref{lastexpresion})
and the constant $\tau$ appearing in the expression of the log likelihood ratio (\ref{lan}) respectively.\\
 \noindent We assume in the problem of testing the two
hypothesis $H_0$ against $H^{(n)}_1$ that the LAN of the the model
(\ref{modelprincipal}) is established, in order to prove the
optimality of the proposed test. To this end, we need the
following assumption :
\begin{enumerate}
    \item[(E.1)] There exists
a $\sqrt{n}$-estimate $\hat{\phi}_n$ of the unknown parameter
$\phi_0$ and a random bounded function $D_n$, such that
\begin{eqnarray*}
 \mathcal{V}_n (\hat{\phi}_{n}) &=& \mathcal{V}_n(\phi_0)- D_n+o_{P}(1).
\end{eqnarray*}
\end{enumerate}
It is now obvious from the previous definitions  that we can state
the following  theorem:\begin{theorem}\label{optimality} Under LAN
and  the conditions (\ref{gradient 1}) (respectively,
(\ref{gradient 2})), $(C.1)$ ($(C.2)$, respectively) and $(E.1)$
the asymptotic power of $\bar{T}_n$ under $H^n_1$ is equal to to
$$1 - \Phi(Z(\alpha)-{\bar{\tau}}^2 ).$$ Furthermore,  $\bar{T}_n$ is
asymptotically optimal.
\end{theorem}
We shall now apply this last theorem in order to conduct
simulations corresponding to the representation of the derived
asymptotic power function. The concerned model is the Nonlinear
time series contiguous to AR(1) processes with an extension to
ARCH processes.
\section{Simulations}\label{section4}
In this section, we assume that $\epsilon_{i}$'s are centered
i.i.d. and $\epsilon_{0} \stackrel{\mathcal{D}}{\longrightarrow}
\mathcal{N}(0,1),$  in this case, we have
$\mathbb{E}({\epsilon_{i}})=0,$ \quad
$\mathbb{E}({\epsilon^2_{i}})=1,$ \quad  and
$\mathbb{E}({\epsilon^4_{i}})=3.$ We treat the case when the
unknown parameter $\phi_0=\rho_0 \in \Theta_1\subset \mathbb{R},$
under  $H_0$, the considering time series model can also
rewritten\begin{eqnarray}
    Y_i =\rho_{0} Y_{i -1} + \epsilon_i\mbox{~~where~~} \quad |\rho_{0}| <1.
\end{eqnarray}

\subsection{Nonlinear time series contiguous to AR(1) processes}
 \noindent To evaluate the performance of our
estimator, we provide   simulations with comment in this section.
In the case when the parameter $\rho_0$ is known, the test $
T_{n}$ is optimal and its power is asymptotically equal to $1 -
\Phi(Z(\alpha)-{\tau}^2 ),$ for more details see \cite[Theorem
3]{HB}. In a general case, when the parameter $\rho_0$ is
unspecified, firstly, we estimate it with the least square
estimates $\hat{\rho}_n= \frac{\sum_{i=1}^{n}Y_i
Y_{i-1}}{\sum_{i=1}^{n} Y^2_{i-1}},$ secondly, with the use of the
(M.E.) under the conditions (\ref{gradient 1}) and $(C.1)$, the
modified estimate $\bar{\rho}_n$ exists and remains
$\sqrt{n}$-consistent, making use of (\ref{perturbation1}) in
connection with the Proposition (\ref{link between the central
sequences in AR1}) it follows:\begin{eqnarray}
  \bar{\rho}_{n} =\frac{D_n}{\dot{\mathcal{V}}_n(\hat{\rho}_n)}
  +\hat{{\rho}}_{n}= \frac{-c_1(\hat{\rho}_n-\rho_0)}{\frac{\dot{\mathcal{V}}_n(\hat{\rho}_n)}{\sqrt{n}}}
  +\hat{{\rho}}_{n}\label{ME},
\end{eqnarray}with the substitution  of the parameter $\rho_0$ by its estimator
$\bar{\rho}_n$ in (\ref{lastexpresion}), we obtain the following
statistics test \begin{eqnarray*}
  \bar{T}_{n}= {\left\{{\frac{\mathcal{V}_{n}( \bar{\rho}_{n})}{\tau( \bar{\rho}_{n})}\geq Z(\alpha)}\right\}}
\mbox{~~where~~~}\bar{\tau}^2=\mathbf{E}({M^2_{f}}({\bar{\epsilon}}_{0,n}))\mathbf{E}(G^2(Y_{0})),\\
\mbox{~and~~~} {\bar{\epsilon}}_{0,n}=Y_0  - Y_{-1}\bar{\rho}_n.
\end{eqnarray*}
It follows from  Theorem (\ref{optimality}) that  $\bar{T}_{n}$
is optimal with an asymptotic  power function  equal to  $1 -
\Phi(Z(\alpha)-{\tau^2(\bar{\rho}_{n})}).$ \\
We choose the function $G$  like this $G: \Big(x_1, x_{2}, \cdot
\cdot \cdot, x_{s},x_{s+1},x_{s+2},\cdot \cdot
\cdot,x_{s+q}\Big)\longrightarrow \frac{5a}{1+x^2_1}
\mbox{~where~} a\neq0.$\\
 In our simulations,  the true
value of the parameter $\rho_0$ is fixed at $0.1$ and the sample
sizes are  fixed at $n = 30, 40, 80$ and $400,$ for a level
$\alpha = 0.05 $, the power relative for each test estimated upon
$m=1000$ replicates, we represent simultaneously the power test
with a true parameter $\rho_0,$ the empirical power test which is
obtained with the replacing the true value $\rho_0$ by its
estimate (M.E.) $\bar{\rho}_{n}$ corresponding to the equality
(\ref{ME}), and the empirical power test which is obtained with
the subsisting the true value $\rho_0$ by its least square
estimator LSE $\hat{\rho_n}$ (an estimator with no correction), we
remark that, the two representations with the true value and the
modified estimate M.E. are close for large $n$.
\begin{figure}[h!]
\includegraphics[scale=0.3]{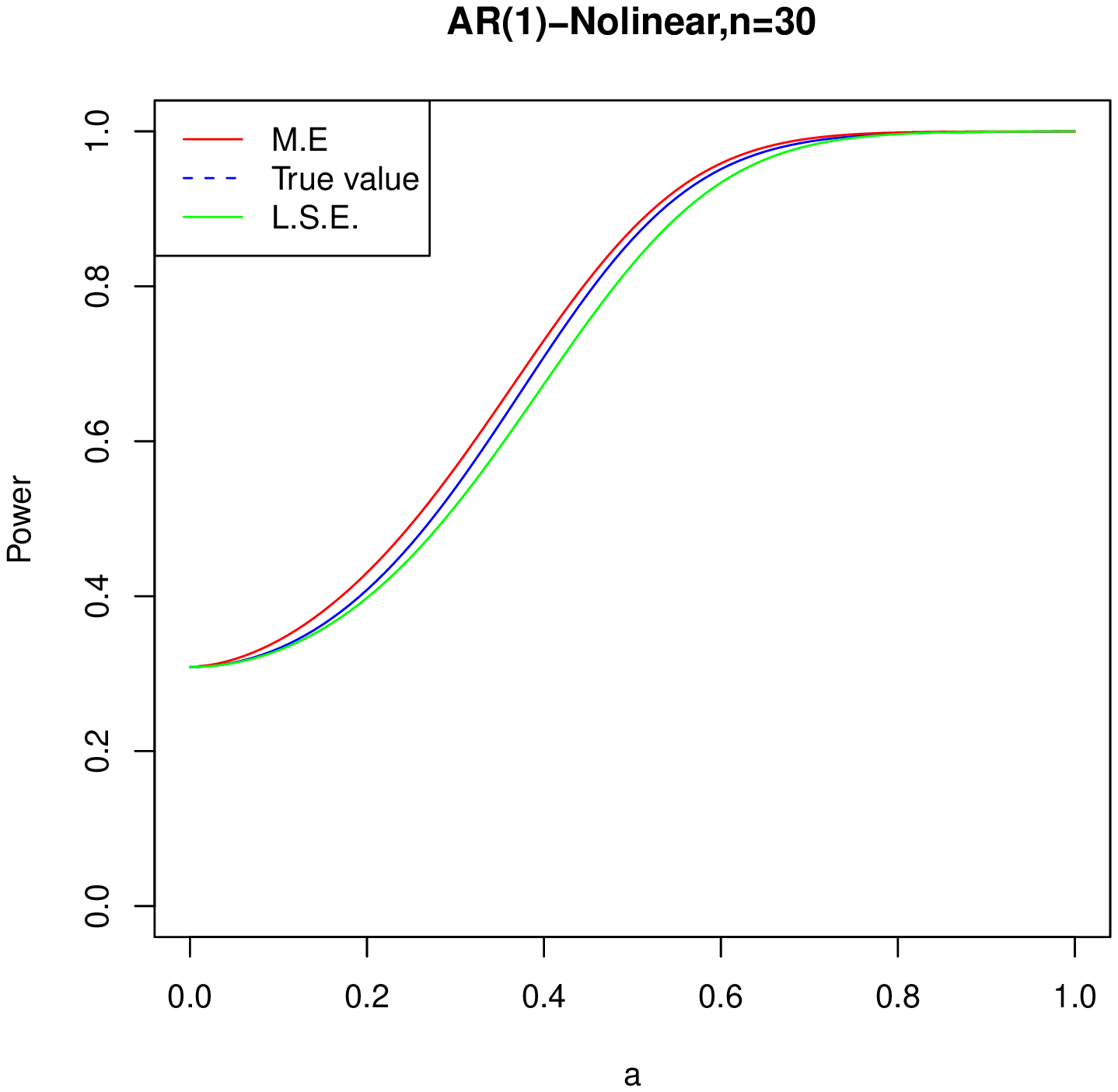}
\includegraphics[scale=0.3]{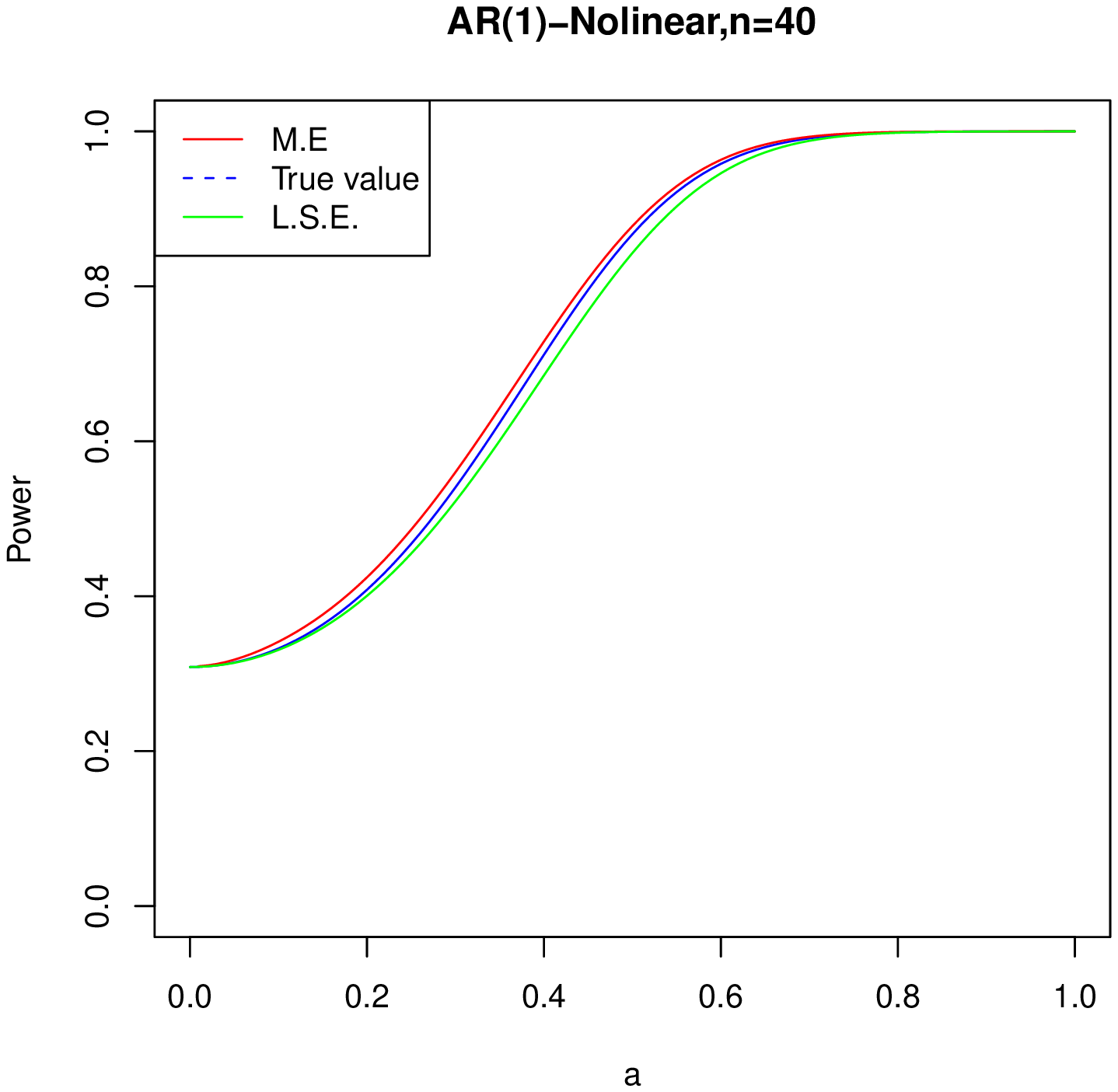}
\includegraphics[scale=0.3]{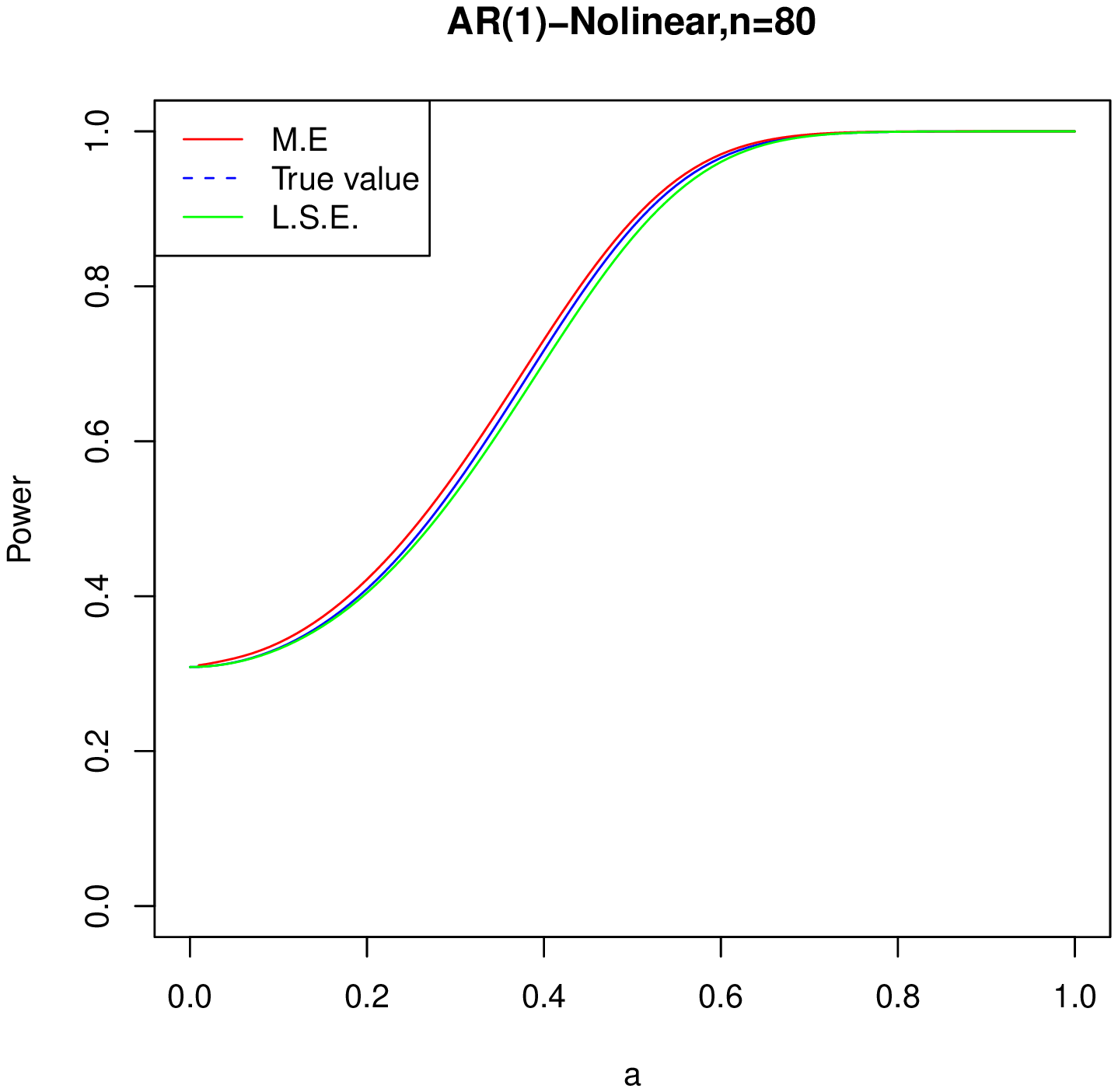}
\includegraphics[scale=0.3]{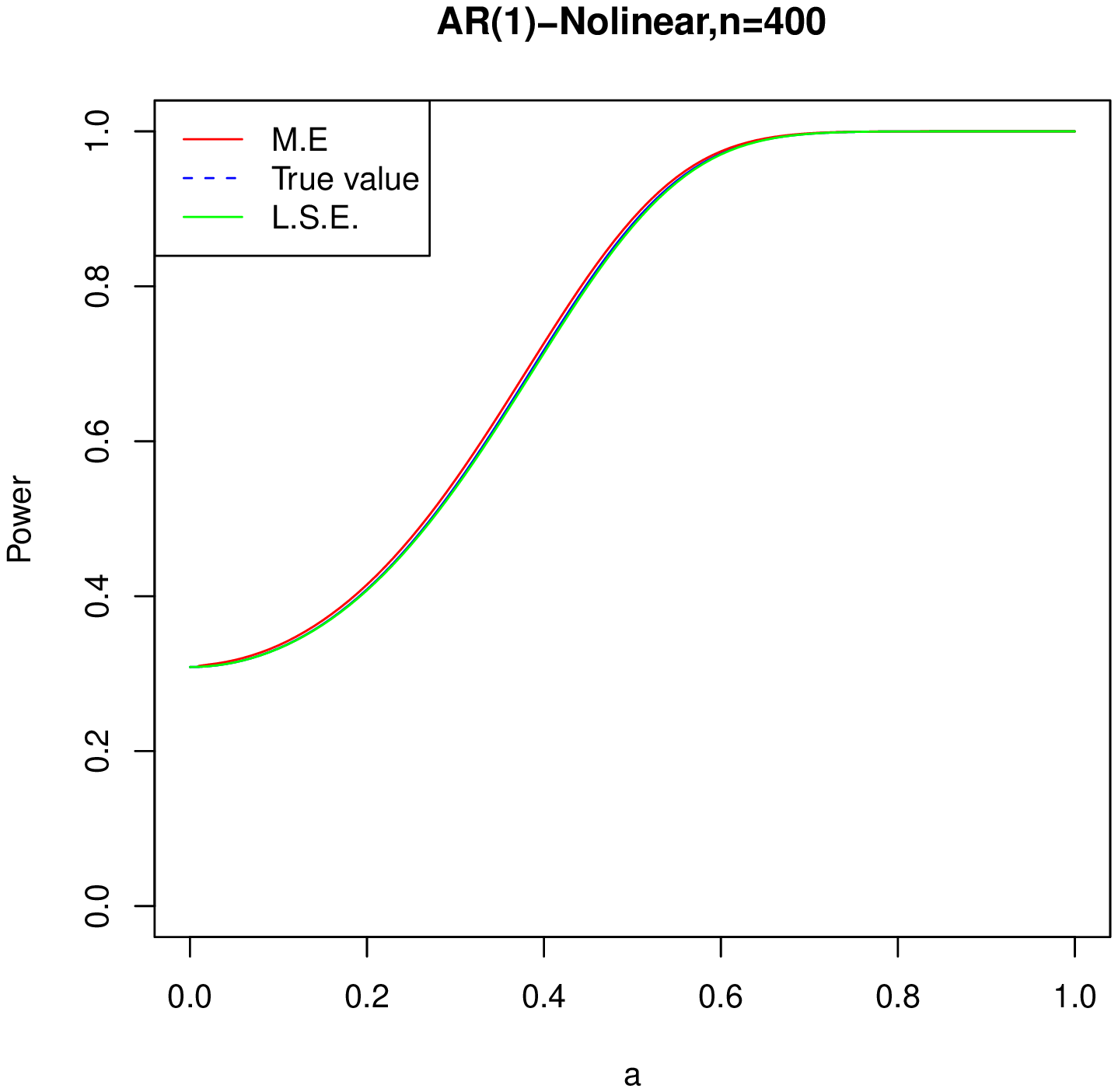}
\end{figure}
\FloatBarrier
\subsection{ ARCH processes}
\noindent  With the substitution of  the parameter $\rho_0$ by its
modified estimate $\bar{\rho}_{n},$ in (\ref{lastexpresion}), we
obtain the following test
 \begin{eqnarray*}
 \bar{ T}_{n} &=& I{\left\{{\frac{\mathcal{V}_{n}(\bar{\rho}_{n})}{\tau(\bar{\rho}_{n})}\geq Z(\alpha)}\right\}},
\end{eqnarray*}
such that\begin{eqnarray*}
  {\bar{\tau}}^2 &=& \bar{I}_{0,n} \mathbf{E}\left(G(Y(0)
        \right)^2
+ \frac{(\bar{I}_{2,n} - 1)}{4} \mathbf{E}\left(B(Y(0) \right)^2 +
\bar{I}_{1,n}
\mathbf{E}\left(G(Y(0))B(Y(0)\right),\\
\bar{I}_{j,n}&=&\mathbf{E}\Big({\bar{\epsilon}}^j_{0,n}{M^2_{f}}({\bar{\epsilon}_{0,n}})\Big)
\mbox{,} \quad j=0,1,2,   \mbox{~~~and~~ }
{\bar{\epsilon}}_{0,n}=Y_0 - Y_{-1} \bar{\rho}_n.
 \end{eqnarray*} In our simulations,  the true value of the parameter  $\rho_0$
is fixed at $ 0.1$ and the sample sizes are fixed at $n = 30, 40,
80$ and $200,$ for a level $\alpha = 0.05 $, the power relative
for each test estimated upon $m=1000$ replicates. We choose the
functions $G$ and $B$ like this $G=B: \Big(x_1, x_{2}, \cdot \cdot
\cdot, x_{s},x_{s+1},x_{s+2},\cdot \cdot
\cdot,x_{s+q}\Big)\longrightarrow \frac{3.5a}{1+x^2_1}
\mbox{~where~} a\neq0.$\\
We represent simultaneously the power test with a true parameter
$\rho_0$ and the empirical power test which is obtained with the
subsisting the true value $\rho_0$ by its estimate (M.E.)
$\bar{\rho}_{n}$ corresponding to the equality (\ref{ME}),we
represent simultaneously the power test with a true parameter
$\rho_0,$ the empirical power test which is obtained with the
subsisting the true value $\rho_0$ by its estimate (M.E.)
$\bar{\rho}_{n}$ corresponding to the equality (\ref{ME}), and the
empirical power test which is obtained with the subsisting the
true value $\rho_0$ by its least square estimator LSE
$\hat{\rho_n}$ (estimator  with no correction), we remark that,
when $n$ is large, we have a similar conclusion as the previous
case .
\begin{figure}[h!]
\includegraphics[scale=0.3]{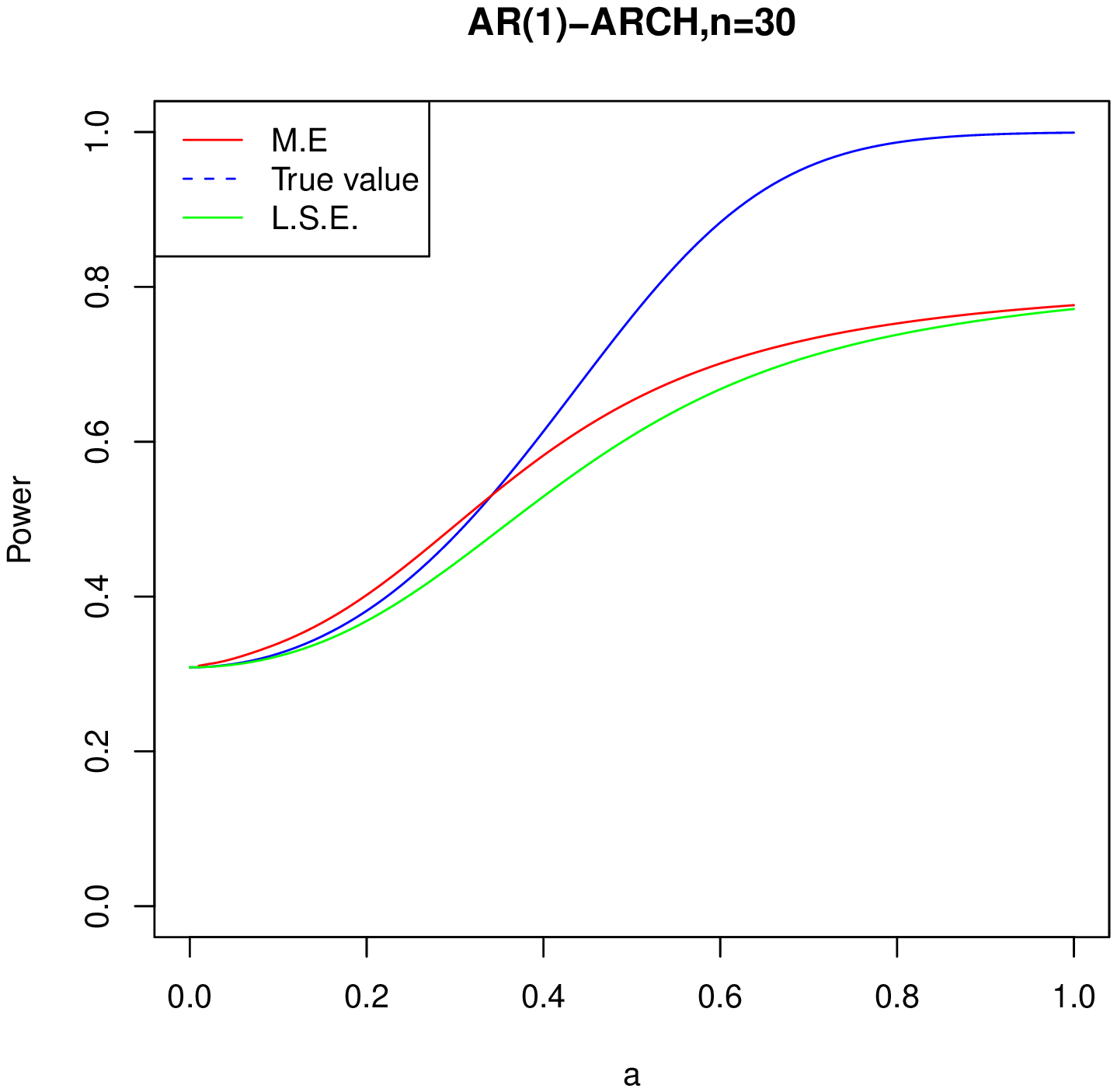}
\includegraphics[scale=0.3]{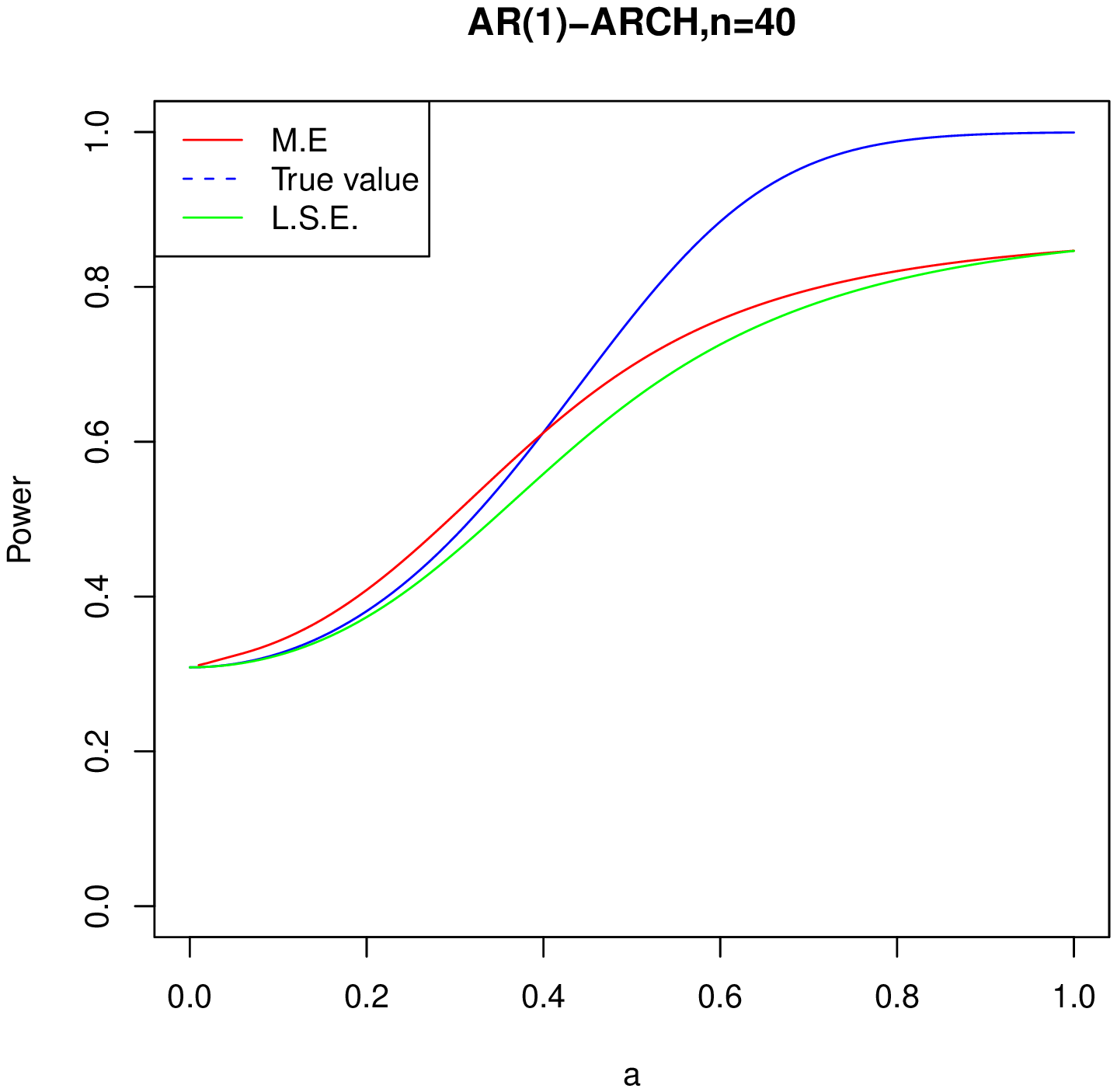}
\includegraphics[scale =0.3]{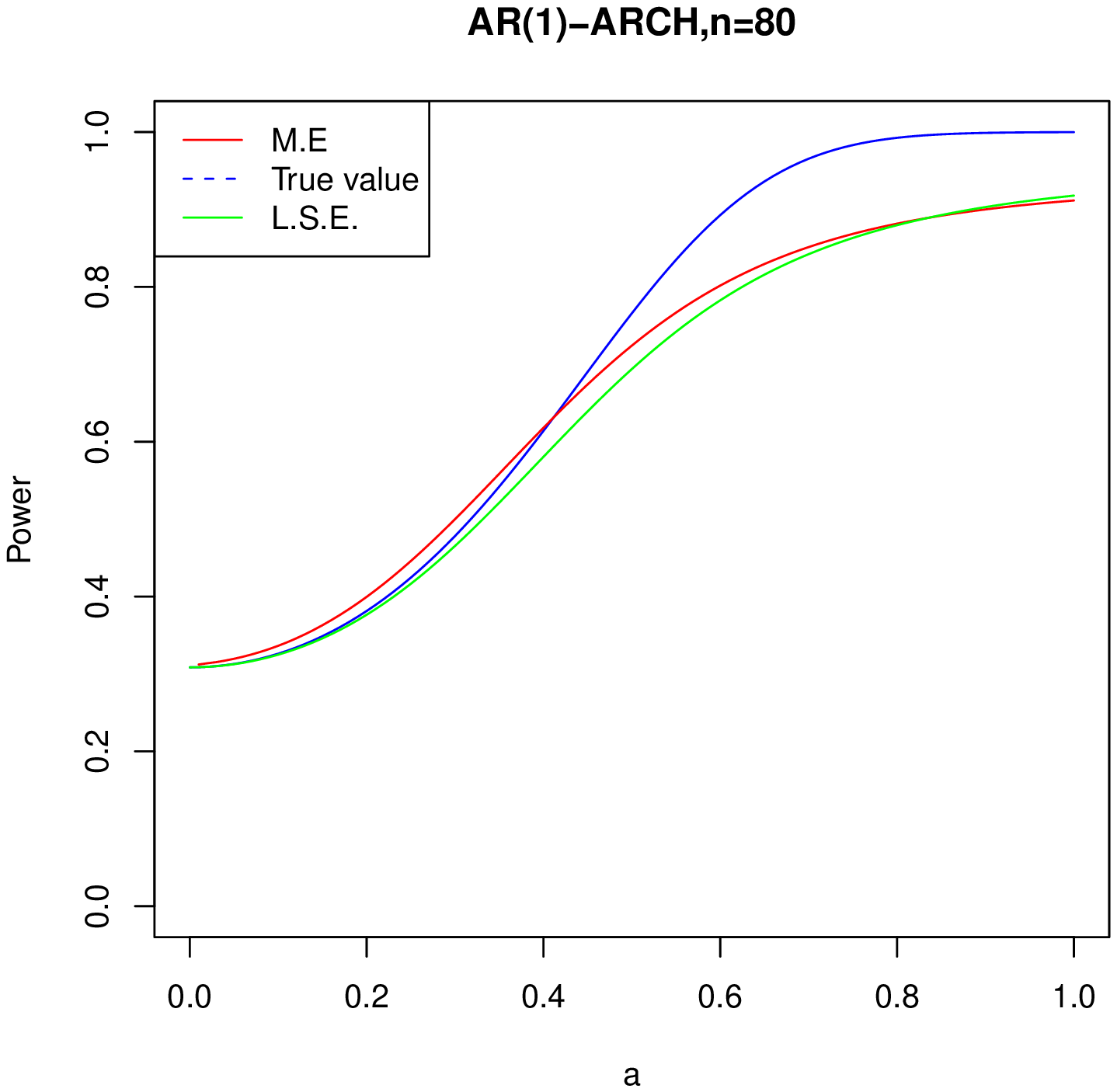}
\includegraphics[scale=0.3]{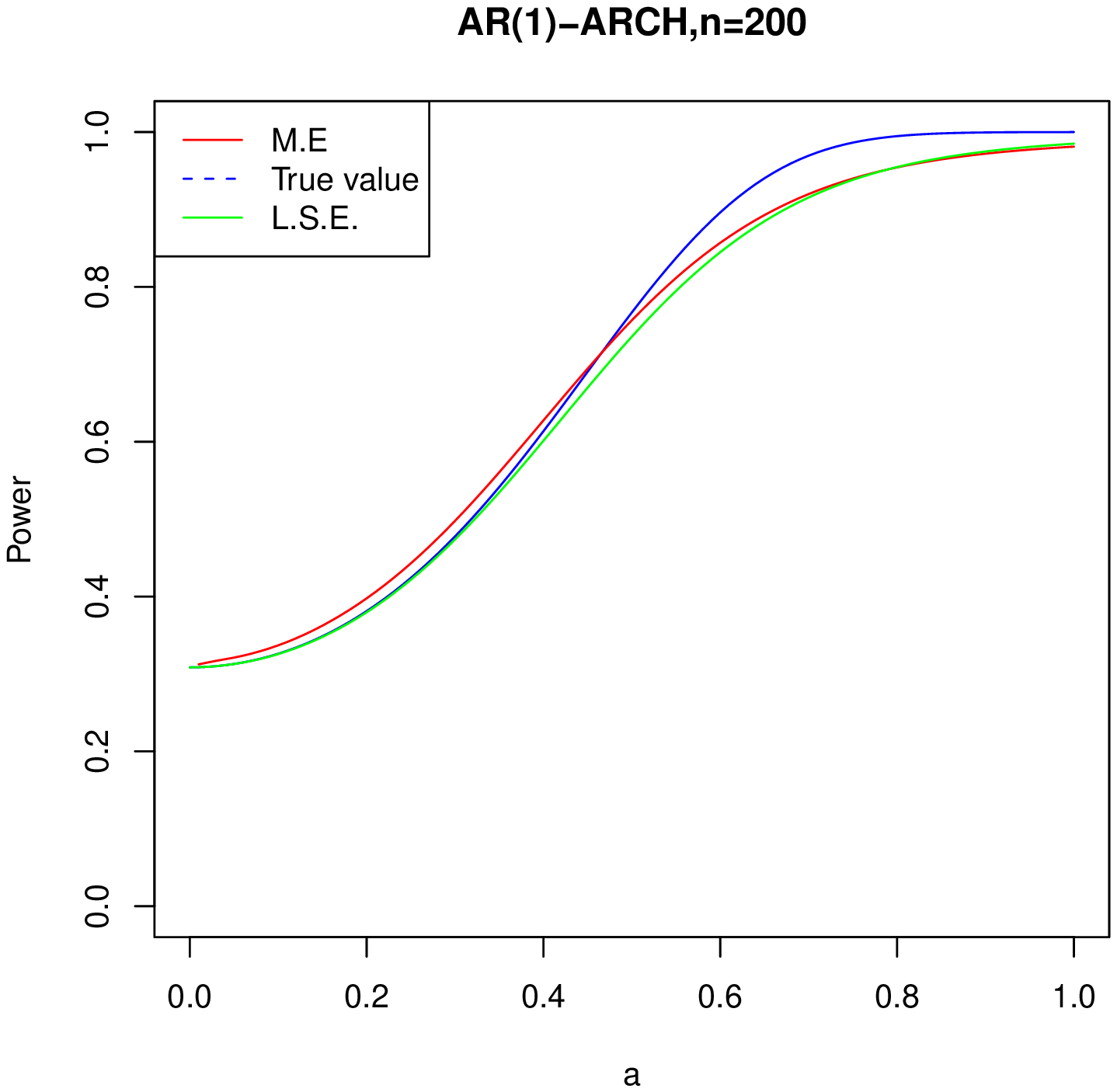}
\end{figure}
 \FloatBarrier

\begin{Remark}
We mention that the limiting distributions appearing in
Proposition (\ref{AR1ECARTfirst}) and Proposition
(\ref{ARCHECARTsecond}) depend on the unknown quantity
$b_n=(\hat{\rho}_n - \rho_0)$, i.e., in practice $\rho_0$ is not
specified, in general. To circumvent this difficulty, we use the
Efron's Bootstrap in order to evaluate $b_n$, more precisely, the
interested reader may refer to the following references :
\cite{Efron1979} for the description of the Bootstrap methods,
\cite{Bertail1994}, \cite{Kvan2007} for the Bootstrap methods in
AR(1) time series models and \cite{Fryzlewicz2008}  for the ARCH
models.
\end{Remark}
\newpage
\section{Proof of the results}\label{proof}
\subsection*{Proof of the Proposition \ref{consistence}}
\noindent Consider the following fundamental
decomposition:\begin{eqnarray}
(\phi_{n}^{(1,j_{n})})^\prime=(\hat{\phi}_{n})^\prime+{(O_{j_{n}})}^\prime,\label{fundamental
decomposition of estimate}\\
\mbox{~~where~~}{O}^\prime_{j_{n}}={(O_{j_{n},i})^\prime}_{i\in\{1,\ldots,{\ell}+p\}},
\mbox{~~such that~~} O_{j_{n},i}=0 \mbox{~~ when ~~}i\neq
j_{n},\nonumber\\
\mbox{~~and~~} O_{j_{n},j_{n}}=\bar{\rho}_{n,j_{n}} -
 \hat{{\rho}}_{n,j_{n}}.\nonumber
 \end{eqnarray}
Firstly, we have
$\hat{\phi}_{n}\stackrel{P}{\longrightarrow}\phi_0$, secondly we
can deduce from (\ref{perturbation1}) that:\begin{eqnarray}
O_{j_{n},j_{n}}=\frac{D_n}{\frac{\partial\mathcal{V}_n(\hat{\phi}_{n})}{\partial\rho_{j_{n}}}}
=\frac{1}{\sqrt{n}}\,D_n\,\frac{1}{{\frac{1}{\sqrt{n}}\,\frac{\partial\mathcal{V}_n(\hat{\phi}_{n})}{\partial\rho_{j_{n}}}}}.\label{dephasage}
\end{eqnarray}Since $D_n$ is bounded, we can remark that
$\frac{1}{\sqrt{n}}\,D_n\stackrel{P}{\longrightarrow}0,$  from
$(C.1)$, there exists some constante $c_1\neq0,$ such that
${\frac{1}{\sqrt{n}}\,\frac{\partial\mathcal{V}_n(\hat{\phi}_{n})}{\partial\rho_{j_{n}}}}\stackrel{P}{\longrightarrow}c_1,$
from (\ref{gradient 1}) and since the function $x
\rightarrow\frac{1}{x}$ is continuous on $\mathbb{R}-\{0\},$ it
follows that the random variable
$\frac{1}{{\frac{1}{\sqrt{n}}\,\frac{\partial\mathcal{V}_n(\hat{\phi}_{n})}{\partial\rho_{j_{n}}}}}\stackrel{P}{\longrightarrow}
\frac{1}{c_1},$ then the couple $\Big(\frac{1}{\sqrt{n}}\,D_n
;\frac{1}{{\frac{1}{\sqrt{n}}\,\frac{\partial\mathcal{V}_n(\hat{\phi}_{n})}{\partial\rho_{j_{n}}}}}
\Big)$ converges in probability to the couple
$\Big(0~;~\frac{1}{c_1}\Big),$ since the function $(x,y)
\rightarrow x y$ is continuous on $\mathbb{R} \times \mathbb{R},$
it result from (\ref{dephasage}), that the random variable
$O_{j_{n},j_{n}}\stackrel{P}{\longrightarrow} \frac{0}{c_1}=0,$
therefore\begin{eqnarray}
{O_{j_{n}}}^\prime=(0,\dots0,O_{j_{n},j_{n}},0\dots0)^\prime
\stackrel{P}{\longrightarrow}
(0,\dots0,0,0\dots0)^\prime.\label{convergence of the error }
\end{eqnarray}

\noindent Consider again the equality (\ref{fundamental
decomposition of estimate}), since the function $(x,y)\rightarrow
x+y$ is continuous on $\mathbb{R}^{{\ell} +p} \times
\mathbb{R}^{{\ell} +p},$  it results  from (\ref{convergence of
the error }) that $\phi_{n}^{(1,j_{n})}$ converges in probability
to $\phi_0$ as\\ $n\rightarrow\infty.$ Notice that the last
previous convergences in probability follow immediately  with the
use of the continuous mapping theorem, for more details, see
\cite{Billingsley1968} or \cite{vandervaart1998}. By following
the same previous reasoning, we shall prove  the consistency of
the estimate $\phi_{n}^{(2,k_{n})}.$
 Note that $\phi_{n}^{(1,j_{n})}$ is $\sqrt{n}$-consistent
 estimate of the parameter $\phi_0$ and $$\sqrt{n}(\phi_{n}^{(1,j_{n})}-\phi_0)=O_{P}(1),$$
 where $O_{P}(1)$ is bounded in probability  in $\mathbb{R}^{{\ell}+p}.$ In fact, it follows from (\ref{fundamental
decomposition of estimate}) that \begin{eqnarray}
\sqrt{n}(\phi_{n}^{(1,j_{n})}-\phi_0)=\sqrt{n}(\hat{\phi}_{n}-\phi_0)
+\sqrt{n} O_{j_{n}}=O_{P}(1)  + \sqrt{n} O_{j_{n}}.\end{eqnarray}
Since
$\sqrt{n}O_{j_{n},j_{n}}=D_n\frac{1}{{\frac{1}{\sqrt{n}}\,\frac{\partial\mathcal{V}_n(\hat{\phi}_{n})}{\partial\rho_{j_{n}}}}}$
 and using the condition $(C.1)$, it results that \\$\sqrt{n}
 O_{j_{n}}=O_{P_{1}}(1),$ where $O_{P_{1}}(1)$ is bounded in probability  in
 $\mathbb{R}.$\\
We deduce
that\begin{eqnarray}\sqrt{n}(\phi_{n}^{(1,j_{n})}-\phi_0)=O_{P}(1).\end{eqnarray}
 Notice that  with a similar argument and with changing $\phi_{n}^{(1,j_{n})}$, $(C.1)$ and (\ref{gradient 1}) by
  $\phi_{n}^{(2,k_{n})}$, $(C.2)$ and (\ref{gradient 2}) respectively, we obtain \begin{eqnarray}\sqrt{n}(\phi_{n}^{(2,k_{n})}-\phi_0)=O_{P}(1).\end{eqnarray}
In order to prove Lemma \ref{sufficientcondition}, we need to stated the following classical lemmas:\\
\begin{lemma}\label{probabilitycomparaison} Let
$(X_i)_{i\in\{1,\dots,l\}}$ be a sequence of a positive random
variables on the probability space $(\Omega,\mathcal{F},P),$
$(\alpha_i)_{i\in\{1,\dots,l\}}$ a sequence of a positive
(strictly) reals such that $\sum_{i=1}^{l}\frac{1}{\alpha_i}=1$,
then we have, for each $\epsilon>0,$
\begin{eqnarray*}
P\left(\sum_{i=1}^{l}X_i>\epsilon\right)&\leq&\sum_{i=1}^{l}P\left(X_i>\frac{\epsilon}{\alpha_i}\right).
\end{eqnarray*}
 \end{lemma}

\begin{lemma}\label{bounded in probability}
Let $(X_n)_{n\geq0}$ be a sequence of a random variables on the
probability space $(\Omega,\mathcal{F},P),$ such that
$X_n=O_{P}(1)$, then  $X^2_n=O_{P}(1)$.
 \end{lemma}

  \subsection*{Proof of the Lemma \ref{probabilitycomparaison}}
\noindent Firstly, we remark that, $\forall \epsilon>0,$  we have
$$\left\{\sum_{i=1}^{l}X_i>\epsilon\right\}\subset\bigcup_{i=1}^{n}\left\{X_i>\frac{\epsilon}{\alpha_i}\right\}.$$
In fact, we suppose there exists
$$\omega\in\left\{\sum_{i=1}^{l}X_i>\epsilon\right\} ~~\mbox{and}~~  \omega\notin
\bigcup_{i=1}^{n} \left\{X_i>\frac{\epsilon}{\alpha_i}\right\},$$
then for each $i \in\{1,\dots,l\},$ we have $X_i(\omega
)\leq\frac{\epsilon}{\alpha_i},$ which implies that
$$\sum_{i=1}^{l}X_i(\omega)\leq\epsilon,$$
hence a contradiction. With the use of the $\sigma$-additivity, we obtain
$$P\left(\sum_{i=1}^{l}X_i>\epsilon\right)\leq P\left(\bigcup_{i=1}^{n}\left\{X_i>\frac{\epsilon}{\alpha_i}\right\}\right)\leq\sum_{i=1}^{l}P\left(X_i>\frac{\epsilon}{\alpha_i}\right).$$

\subsection*{Proof of the Lemma \ref{sufficientcondition}}
\noindent In this case $\phi_0=\rho_0 \in \Theta _1 \subset
\mathbb{R},$ we denote by $\hat{\rho}_n$ the $\sqrt{n}$-consistent estimator of $\rho_0.$  \\
 Let $A>0$, from the triangle inequality combined with
the Lemma (\ref{probabilitycomparaison}),we obtain:
\begin{eqnarray*}
 \lefteqn{ P\left(\left|\frac{1}{\sqrt{n}}
{\mathcal{\dot{V}}}_n(\hat{\rho}_n)-c_1\right|>A\right)}\\&=&
P\left(\left|\frac{1}{\sqrt{n}}
{\mathcal{\dot{V}}}_n(\hat{\rho}_n)-\frac{1}{\sqrt{n}}
{\mathcal{\dot{V}}}_n(\rho_0)\right|+\left|\frac{1}{\sqrt{n}}
{\mathcal{\dot{V}}}_n(\rho_0)-c_1\right|>A\right)\\
 &\leq&P\left(\left|\frac{1}{\sqrt{n}}
{\mathcal{\dot{V}}}_n(\hat{\rho}_n)-\frac{1}{\sqrt{n}}
{\mathcal{\dot{V}}}_n(\rho_0)\right|>\frac{A}{2}\right)+P\left(\left|\frac{1}{\sqrt{n}}
{\mathcal{\dot{V}}}_n(\rho_0)-c_1\right|>\frac{A}{2}\right).
\end{eqnarray*}
Firstly, we have \begin{eqnarray}
  P\left(\left|\frac{1}{\sqrt{n}}
{\mathcal{\dot{V}}}_n(\rho_0)-c_1\right|>\frac{A}{2}\right)\rightarrow0
\mbox{~~as~~}n\rightarrow\infty, \label{firststeep}
\end{eqnarray}
Secondly, we have \begin{eqnarray} \left|\frac{1}{\sqrt{n}}
{\mathcal{\dot{V}}}_n(\hat{\rho}_n)-\frac{1}{\sqrt{n}}
{\mathcal{\dot{V}}}_n(\rho_0)\right|= \frac{1}{\sqrt{n}}\left|
{\mathcal{\ddot{V}}}_n(\tilde{\rho}_n)\right|\Big|\hat{\rho}_n
-\rho_0 \Big|\\\label{boundedconssitency second derivative}
 =\frac{1}{\sqrt{n}}\left|\frac{1}{\sqrt{n}}
{\mathcal{\ddot{V}}}_n(\tilde{\rho}_n)
\right|\Big|\sqrt{n}(\hat{\rho}_n -\rho_0)
\Big|,\label{boundedconssitency second derivative}
\end{eqnarray}
where $\tilde{\rho}_n$ is a point between $\rho_0$ and
$\hat{\rho}_n,$ then there exists a sequence $\eta_n$ with values
in the interval $[0,1]$, such that $\tilde{\rho}_n = \eta_n \rho_0
+ (1-\eta_n)\hat{\rho}_n,$ this implies that \\$|\tilde{\rho}_n
-\rho_0 |\leq(1-\eta_n)|\hat{\rho}_n - \rho_0|\leq |\hat{\rho}_n -
\rho_0|,$ this last inequality enable us to concluded that
$\tilde{\rho}_n$ is $ \sqrt{n}$-consistency estimator of $\rho_0$,
it follows from $(C.3)$ applied on the equality
(\ref{boundedconssitency second derivative}) that\begin{eqnarray}
P\left(\left|\frac{1}{\sqrt{n}}
{\mathcal{\dot{V}}}_n(\hat{\rho}_n)-\frac{1}{\sqrt{n}}
{\mathcal{\dot{V}}}_n(\rho_0)\right|>\frac{A}{2}\right)\rightarrow
0 \mbox{~~as~~} n\rightarrow 0.\end{eqnarray} Thus we obtain
$(i).$

\subsection*{Proof of Proposition \ref{M.E}} \noindent It suffices to choose under
(\ref{gradient 1}) and $(C.1)$ the estimate
$\bar{\phi_{n}}=\phi_{n}^{(1,j_{n})},$ or under (\ref{gradient 2})
and $(C.2)$ the estimate $\bar{\phi_{n}}=\phi_{n}^{(2,k_{n})}.$

In order to prove the  Proposition (\ref{AR1ECARTfirst}), we need
a following classical result.
\begin{lemma}\label{lemme convergence in probability}
Let $(\Omega,\mathcal{F},P)$ be a probability space,
$(X_{n})_{n\geq 1}$ is a sequence of real random variables on
$\Omega$. If $X_{n}$ converges in probability to  a constant $c$,
then,
 there exists a sequence of random variable  $(Y_{n})_{n}$,
with   $X_{n}= c + Y_{n}$, such that, $Y_{n}$ converges in
probability to $0$.
\end{lemma}

\subsection*{Proof of Lemma \ref{lemme convergence in probability}}
\noindent For all $A>0$, we have:
$$P\left(|Y_{n}|>A\right)=P\left(|X_{n}- c|>A\right)\rightarrow 0, ~~\mbox{as}~~
n\rightarrow \infty.$$

\subsection*{Proof of Lemma \ref{bounded in probability}}
For all $\epsilon>0,$ $\exists  M_1>0$ such that:\\
$\sup_{\alpha}\Big((P\left(|X_{\alpha}|>M_1\right)\Big)<\epsilon,$
this  implies that
$\sup_{\alpha}\Big((P\left(|X_{\alpha}|^2>M^2_1\right)\Big)<\epsilon,$
therefore with the choice of $M=M_1$, we obtain the result.

\subsection*{Proof of Proposition \ref{AR1ECARTfirst}}
 $\epsilon_{i}$'s are centered
i.i.d. and $\epsilon_{0} \stackrel{\mathcal{D}}{\longrightarrow}
\mathcal{N}(0,1),$ making use of  the results of \cite[Theorem
2]{HB}, we have
\begin{eqnarray*}
\mathcal{V}_{n}(\rho_0)=-\frac{1}{\sqrt{n}}\sum_{i=1}^{n}M_{f}(\epsilon_i)G(Y(i-1)).
\end{eqnarray*}
The estimated central sequence is
$$\mathcal{V}_{n}(\hat{\rho}_n)=-\frac{1}{\sqrt{n}}\sum_{i=1}^{n}M_{f}(\hat{\epsilon}_{i,n})G(Y(i-1)).$$
By Taylor expansion with order $2,$ we have :
\begin{eqnarray}
\mathcal{V}_{n}(\hat{\rho}_n) -
\mathcal{V}_{n}(\rho_0)=\dot{\mathcal{V}_{n}}(\hat{\rho}_n)(\hat{\rho}_n
- \rho_0) +
\frac{1}{2}\mathcal{\ddot{V}}_{n}(\tilde{\rho_{n}})(\hat{\rho}_n -
\rho_0)^2 \label{boundedcsexample1},
\end{eqnarray}
where $\tilde{\rho_{n}}$ is a point between $\rho_0$ and
$\hat{\rho}_n$ and \begin{eqnarray*}
\dot{\mathcal{V}_{n}}(\tilde{\rho_{n}})=\frac{-1}{\sqrt{n}}\sum_{i=1}^{n}
Y_{i-1}G(Y(i-1)).
\end{eqnarray*}
Note that  \begin{eqnarray*}
R_{n}=\frac{1}{2}\mathcal{\ddot{V}}_{n}(\tilde{\rho_{n}})(\hat{\rho}_n-\rho_0)^2=
\frac{1}{2\sqrt{n}}
\frac{1}{\sqrt{n}}\mathcal{\ddot{V}}_{n}(\tilde{\rho_{n}})\Big(\sqrt{n}
(\hat{\rho}_n-\rho_0)\Big)^2.
\end{eqnarray*}
Since the estimator $\hat{\rho}_{n}$ is $\sqrt{n}$-consistent and
with the use of Lemma (\ref{bounded in probability}), it results
that $$\Big(\sqrt{n} (\hat{\rho}_n-\rho_0)\Big)^2=O_{P}(1),$$ from
the assumption $(C.3),$ it follows that $$R_{n} = o_{P}(1),$$
finally we deduce that,
\begin{eqnarray} \mathcal{V}_{n}(\hat{\rho}_n) -
\mathcal{V}_{n}(\rho_0)=\dot{\mathcal{V}_{n}}(\hat{\rho}_n)(\hat{\rho}_n
- \rho_0) + o_{P}(1)\label{firstboundedEQUALITY}.
\end{eqnarray}
This implies that\begin{eqnarray}
\frac{\dot{\mathcal{V}_{n}}(\hat{\rho}_n)}{\sqrt{n}}-\frac{\dot{\mathcal{V}_{n}}(\rho_0)}{\sqrt{n}}=
\frac{\mathcal{\ddot{V}}_n(\check{\rho_{n}})}{\sqrt{n}}
(\hat{\rho}_n-\rho_0) +
o_{P}(1)=\frac{1}{\sqrt{n}}\frac{{\mathcal{\ddot{V}}_{n}}
(\check{\rho_{n}})}{\sqrt{n}} \sqrt{n}(\hat{\rho}_n - \rho_0) +
o_{P}(1),\nonumber\\
\end{eqnarray}where $\check{\rho_{n}}$ is between $\hat{\rho}_n$ and $\rho_0,$
and  $\mathcal{\ddot{V}}_n$ is the second derivative of
$\mathcal{V}_{n}.$ From the assumption $(C.3)$, we have
$$\frac{1}{\sqrt{n}}\frac{\mathcal{\ddot{V}}_n(\check{\rho}_n)}{\sqrt{n}}=o_{P}(1),$$ since the estimator
$\hat{\rho}_{n}$ is $\sqrt{n}$-consistent, it result that
\begin{eqnarray*}
\frac{\dot{\mathcal{V}_{n}}(\hat{\rho}_n)}{\sqrt{n}}-\frac{\dot{\mathcal{V}_{n}}(\rho_0)}{\sqrt{n}}=o_{P}(1),
\end{eqnarray*} this implies that\begin{eqnarray}
\frac{\dot{\mathcal{V}_{n}}(\hat{\rho}_n)}{\sqrt{n}}=\frac{\dot{\mathcal{V}_{n}}(\rho_0)}{\sqrt{n}}
+ o_{P}(1).\label{sqrtsequences}
\end{eqnarray}
With the use of (\ref{sqrtsequences}), the equality
(\ref{firstboundedEQUALITY}) can also rewritten\begin{eqnarray}
\mathcal{V}_{n}(\hat{\rho}_n) -
\mathcal{V}_{n}(\rho_0)=\frac{\mathcal{\dot{V}}_{n}(\hat{\rho}_n)}{\sqrt{n}}
\sqrt{n}(\hat{\rho}_n - \rho_0) + o_{P}(1),\nonumber\\
=\frac{\mathcal{\dot{V}}_{n}(\rho_0)}{\sqrt{n}}
\sqrt{n}(\hat{\rho}_n - \rho_0) +
o_{P}(1)\label{lastconsistencyerodictheorem}.
\end{eqnarray}It follows from the assumption $(C.1)$ combined with the
ergodicity and the stationarity of the model that, the random
variable $\frac{1}{\sqrt{n}}\dot{\mathcal{V}_{n}}(\rho_{0})$
converges in probability to the constant $c_1$, as $n\rightarrow
+\infty$, where
$$c_1=-\mathbb{E}\Big[Y_{0}G(Y(0))\Big],$$ therefore from the Lemma
(\ref{lemme convergence in probability}), there exists a random
variable $X_n,$ $ X_n \stackrel{P}{\longrightarrow}0$ such that
\begin{eqnarray*}
\frac{1}{\sqrt{n}}\dot{\mathcal{V}_{n}}(\rho_0)=c_1
+X_n.\end{eqnarray*} We deduce from the equality
(\ref{lastconsistencyerodictheorem}) and the
$\sqrt{n}$-consistence of the estimator $\hat{\rho_{n}}$, that
\begin{eqnarray}
\mathcal{V}_{n}(\hat{\rho}_n)-\mathcal{V}_{n}(\rho_0)=c_1\sqrt{n}(\hat{\rho}_n
- \rho_0) + o_{P}(1)= -D_n + o_{P}(1),\label{boundedcsexample3}
\end{eqnarray} where $D_n=-c_1 \sqrt{n}(\hat{\rho}_n- \rho_0).$
\noindent Recall that the second derivative
$\mathcal{\ddot{V}}_{n}$ is equal to $0,$ this implies that the
assumption $(C.3)$ is satisfied.

\subsection*{Proof of Proposition \ref{ARCHECARTsecond}}
\noindent The assumption  $(C.1)$ remains satisfied  and the proof
is similar as the proof of Proposition (\ref{AR1ECARTfirst}), in
this case, for all $\rho \in \Theta_1,$ we have\begin{eqnarray*}
\mathcal{\ddot{V}}_{n}(\rho)=\frac{-1}{\sqrt{n}}\sum_{i=1}^{n}
Y^2_{i-1}B(Y(i-1)) 2 \dot{M}_f(\rho).
\end{eqnarray*}
By a simple calculus and since the the function $f$ is the density
of the standard normal distribution, it is easy to prove that the
quantity $ 2\dot{M}_f(\rho)$ is bounded, therefore, there exists a
positive constant $w$ such that $ 2\dot{M}_f(\rho)\leq w$, then
\begin{eqnarray*}
|\frac{1}{\sqrt{n}}\mathcal{\ddot{V}}_{n}(\rho)|\leq
 w\frac{1}{n}\sum_{i=1}^{n} Y^2_{i-1}|B(Y(i-1))|.
\end{eqnarray*}
With the choice $B(Y(i-1))=\frac{2a}{1+Y^2_{i-1}}$ with $a\neq0$,
it results that
\begin{eqnarray*}
|\frac{1}{\sqrt{n}}\mathcal{\ddot{V}}_{n}(\rho)|\leq
 2w |a| \frac{1}{n}\sum_{i=1}^{n} Y^2_{i-1}.
\end{eqnarray*}
By the use of the ergodicity of the model and since the model is
with finite second moments, it follows that the random variable
$\frac{1}{n}\sum_{i=1}^{n}
Y^2_{i-1}\stackrel{a.s}{\longrightarrow}k,$ where $k$ is some
constant, this implies that the condition $(C.3)$ is
straightforward .

\subsection*{Proof of the Theorem \ref{optimality}}
\noindent From the conditions (\ref{gradient 1}) ((\ref{gradient
2}), respectively), $(C.1)$ ($(C.2)$,  respectively), it results
the existence and the $\sqrt{n}$-consistency of the modified
estimate estimate $\bar{\phi}_n$ corresponding to the equation
(\ref{perturbation1}) ((\ref{perturbation2}), respectively). The
combinaison of the condition $(E_1)$ and the Proposition
(\ref{M.E}) enable us to get under $H_0$ the following
equality\begin{eqnarray*}
 \mathcal{V}_n(\bar{\phi}_{n}) &=&
 \mathcal{V}_n(\phi_0)+o_{P}(1).
\end{eqnarray*}This last equation implies that with $o_{P}(1),$ the estimate
central and central sequences are equivalent, in the expression of
the test (\ref{test}), the replacing of the central sequence by
the estimate central sequence has no effect. LAN implies the
contiguity of the two hypothesis (see, \cite[Corrolary
4.3]{Droesbeke}), by Le Cam third lemma's (see for instance,
\cite[Theorem 2]{HM}), under $H^{(n)}_1$, we have $$\mathcal{V}_n
\stackrel{\mathcal{D}}{\longrightarrow}
\mathcal{N}({\tau}^2,{\tau}^2).$$ It follows from the convergence
in probability of  the estimate $\bar{\phi}_n$ to $\phi_0$, the
continuity of the function $\tau : \cdot \longrightarrow
\tau(\cdot)$ and the application of the continuous mapping theorem
 see, for instance (\cite{vandervaart1998}) or \cite{Billingsley1968}, that asymptotically, the power of the
test is not effected when we replace the unspecified parameter
$\phi_0$ by it's estimate, $\bar{\phi}_n,$ hence the optimality of
the test. The power function of the test is asymptotically equal
to $1 - \Phi(Z(\alpha)-{\tau}^2(\bar{\phi}_n)),$ the proof is
similar as
 \cite[Theorem 3]{HB}.

\end{document}